\documentclass[graphics,floatfix, nofootinbib,tightenlines,nobibnotes, 12pt]{revtex4-1}
\usepackage{graphicx}
\usepackage[english]{babel}
\usepackage[utf8]{inputenc}
\usepackage{amsmath,amssymb}
\usepackage{amsfonts}
\usepackage{multirow}
\usepackage{pstricks}
\usepackage{float}
\usepackage{subfigure}
\usepackage{color}
\usepackage{epsfig}
\usepackage{slashed}
\usepackage[colorlinks=true, linkcolor=blue, citecolor=blue, urlcolor=blue]{hyperref}
\usepackage[title]{appendix}

\begin{document}
\title{Excited doubly heavy baryons production via Higgs decays}

\author{Hong-Hao Ma$^{1,2}$}
\email{mahonghao@pku.edu.cn}
\author{Juan-Juan Niu$^{2,3}$}
\email{niujj@gxnu.edu.cn, corresponding author}

\address{$^{1}$ Center for High Energy Physics, Peking University, Beijing 100871, China}
\address{$^{2}$ Guangxi Key Laboratory of Nuclear Physics and Technology, Guangxi Normal University, Guilin 541004, China}
\address{$^{3}$ Department of Physics, Guangxi Normal University, Guilin 541004, China}

\date{\today}

\begin{abstract}
Through the interaction of Higgs with heavy quarks in standard model, we have systematically studied and predicted the production of excited doubly heavy baryons based on non-relativistic QCD theory. 
The decay widths, differential distributions, and major theoretical uncertainties of the excited doubly heavy baryons via the process $H \rightarrow \langle QQ'\rangle[n] \rightarrow \Xi_{QQ'}+ \bar {Q'} + \bar {Q}$ are discussed in detail. The spin and color quantum number of the intermediate $P$-wave diquark state $\langle QQ'\rangle[n]$ can be $\langle cc\rangle[^{1}P_{1}]_{\mathbf{\bar 3}}$, $\langle cc\rangle[^{3}P_{J}]_{\mathbf{6}}$, 
$\langle bc\rangle[^{1}P_{1}]_{\mathbf{\bar 3}/ \mathbf{6}}$, $\langle bc\rangle[^{3}P_{J}]_{\mathbf{\bar 3}/ \mathbf{6}}$, 
$\langle bb \rangle[^{1}P_{1}]_{\mathbf{\bar 3}}$ and $\langle bb\rangle[^{3}P_{J}]_{\mathbf{6}}$, with $J=0, 1, 2$. The contributions from all summed $P$-wave states can be about $3.05\%$, $3.23\%$ and $2.19\%$ of the $S$-wave states for the production of $\Xi_{cc}$, $\Xi_{bc}$ and $\Xi_{bb}$, accordingly. Therefore, there will be about 0.41$\times10^4$ events of $\Xi_{cc}$, 6.35$\times10^4$ events of $\Xi_{bc}$ and 0.28$\times10^4$ events of $\Xi_{bb}$ produced per year at the HL-LHC, and a smaller number of events would be produced at the CEPC or ILC but with a cleaner background to be measured by the experiments.


\end{abstract}

\maketitle

\section{Introduction}

According to the Higgs mechanism of the standard model (SM) in particle physics, Higgs boson is considered as the mass origin of many elementary particles, including fermions and gauge bosons, and the coupling strength of Higgs and fermioms (gauge bosons) is proportional to the corresponding fermion mass (gauge boson squared mass). 
The properties of Higgs themselves and their interactions with heavy quarks and gauge bosons require further investigation. After ten years of theoretical and experimental research since the Higgs scalar boson was first discovered at the Large Hadron Collider (LHC) in 2012, its properties are almost consistent with those predicted by the SM. 
The decay channels of Higgs have been measured separately by ATLAS and CMS collaborations, with similar results~\cite{ATLAS:2022vkf, CMS:2022dwd}, which is also consistent with what the SM predicted.

The largest contribution to the production of doubly heavy baryons comes from the directly hadronic production mechanism~\cite{Chang:2006eu,Chang:2006xp,Berezhnoy:1998aa}, including the gluon-gluon fusion mechanism and the intrinsic charm/bottom mechanisms. However, studying the indirect production mechanism of doubly heavy baryons through Higgs boson decay provides a new platform for further study of Higgs properties in current and future new ``Higgs factories".
Since the electroweak interaction is generally weaker than the strong interaction, the contribution for the production of doubly heavy baryon through the preliminary decay channel $H \rightarrow Z^{0}Z^{0}$ is small enough to be ignored. As for the decay channel $H \rightarrow gg \rightarrow \Xi_{QQ^{\prime}}+ X$, its contribution is also small for the suppression of fermion loop and higher order of strong coupling constant ($\alpha^{4}_{s}$) for the leading-order calculation, where $Q^{(\prime)}$ is labeled as the heavy $c$ or $b$ quark throughout the paper corresponding to the production of $\Xi_{cc}$, $\Xi_{bc}$ and $\Xi_{bb}$. These claims have been tested numerically in our previous paper~\cite{Niu:2019xuq}, to be specific, the decay widths through $H \rightarrow Z^0Z^0/gg$ channels are only a few percent compared to that through $H \rightarrow Q\bar{Q} / Q^{\prime} \bar{Q^{\prime}}$. The initial decay mode of Higgs into the quark-antiquark pairs provides the dominant contribution to the production of doubly heavy baryons, which is at the $\alpha^{2}_{s}-$order level. And this decay process, $H \rightarrow Q\bar{Q}/Q^{\prime}\bar{Q^{\prime}} \rightarrow \Xi_{QQ^{\prime}} + \bar {Q^{\prime}} + \bar{Q}$, is also helpful for studying the Yukawa coupling.

Quark models~\cite{GellMann:1964nj,Zweig:1981pd,Zweig:1964jf} predicted the existence of doubly heavy baryons, such as $\Xi^{++}_{cc}$, $\Xi^{+}_{cc}$, $\Omega^{+}_{cc}$, $\Xi^{+}_{bc}$, $\Xi^{0}_{bc}$, $\Omega^{0}_{bc}$, $\Xi^{0}_{bb}$, $\Xi^{-}_{bb}$ and $\Omega^{-}_{bb}$, but it wasn't until 2017 that the first doubly heavy baryon $\Xi_{cc}^{++}$ was experimentally detected by the LHCb collaboration~\cite{Aaij:2017ueg}. In this paper, we will not consider the isospin-breaking effect and, for convenience, use $\Xi_{cc}$ for undistinguished $\Xi^{++}_{cc}$, $\Xi^{+}_{cc}$, and $\Omega^{+}_{cc}$, $\Xi_{bc}$ for undistinguished $\Xi^{+}_{bc}$, $\Xi^{0}_{bc}$, and $\Omega^{0}_{bc}$, and $\Xi_{bb}$ for undistinguished $\Xi^{0}_{bb}$, $\Xi^{-}_{bb}$ and $\Omega^{-}_{bb}$. Searching for other undetected doubly heavy baryons remains one of the important topics of the LHCb experiment. Theoretically, many preliminary researches on both the direct and indirect production mechanism of doubly heavy baryons are predicted~\cite{Baranov:1995rc,Ma:2003zk,Li:2007vy,Chen:2014frw,Huan-Yu:2017emk,Baranov:1995rc,Doncheski:1995ye,Zheng:2015ixa,Jiang:2012jt,Yao:2018zze,Chen:2018koh,Niu:2018ycb,Berezhnoy:1998aa}, which shall be helpful to guide the development of experiments and be a verification of the quark model. Some previous studies have been carried out through the Higgs decay into hadrons, such as $B_c$, $J/\psi$, $\Upsilon$, $\Xi_{QQ^{\prime}}$, and so on~\cite{Bodwin:2013gca,Bodwin:2014bpa,Qiao:1998kv,Jiang:2015pah,Liao:2018nab,Aad:2015sda}.

Non-relativistic Quantum Chromodynamics (NRQCD)~\cite{Bodwin:1994jh,Petrelli:1997ge}, fragmentation function method~\cite{Falk:1993gb,Doncheski:1995ye,Ma:2003zk} and phenomenological model~\cite{Ali:2018ifm,Ali:2018xfq,Qin:2020zlg} are all very effective theoretical tools for calculating the doubly heavy baryons production. In this paper, we would theoretically discuss the production of excited doubly heavy baryons through indirectly Higgs decays in the NRQCD framework. The production of doubly heavy baryons can be factorized into two parts. One part is the non-perturbative long-distance matrix elements, which represents the transition probability from the intermediate diquark state $\langle QQ^{\prime} \rangle$ binding into the corresponding doubly heavy baryons $\Xi_{QQ^{\prime}}$. Meanwhile the transition probability can be approximatively related to the Schr\"{o}dinger wave function at the origin $|\Psi_{QQ^{\prime}}(0)|$ for $S$-wave states or the first derivative of the wave function at the origin $|\Psi^{\prime}_{QQ^{\prime}}(0)|$ for $P$-wave states, which can be obtained by fitting experimental data or some non-perturbative methods. Another is the short-distance coefficient, which can be calculated perterbatively, and we mainly focus our attention on this part. It is worth emphasizing that the decay widths via process $H \rightarrow Z^{0}Z^{0} / gg \rightarrow \Xi_{QQ^{\prime}} + X $, are so small and will not be considered here. For the $\Xi_{cc}$ production, the intermediate decay channel $H \rightarrow c\bar{c}$ dominates. As for the production of $\Xi_{bc}$ and $\Xi_{bb}$, the dominant decay channel is $H^0 \rightarrow b\bar{b}$, which is the largest decay channel of Higgs with a branching ratio of 58$\%$~\cite{Patrignani:2016xqp,Niu:2018otv}.

At the LHC, the most important production mechanism of Higgs bosons is gluon-gluon fusion (ggF) and through this production mechanism, a large number of Higgs particles would be produced at the High Luminosity LHC (HL-LHC)~\cite{Bediaga:2012py,Bediaga:2018lhg,LHCHIGGS}, or even a upgraded High Energy LHC (HE-LHC). At the $e^{+}e^{-}$ colliders, the main Higgs production mechanism is Higgs and $Z^{0}$ asociated production. More than one million Higgs events would be produced at the Circular Electron Positron Collider (CEPC)~\cite{CEPCStudyGroup:2018rmc} and the International Linear Collider (ILC)~\cite{Simon:2012ik}. Thus, HL/HE-LHC and CEPC/ILC provide a good experimental platform for studying the indirect production mechanism of excited doubly heavy baryons via Higgs decays. A detailed study of the excited doubly heavy baryons shall supplement the study of ground states baryons and determine the contribution of $P$-wave states.

The remaining parts of this paper are organized as follows. In Sec.II, we introduce the calculation technology for the production of excited doubly heavy baryons through Higgs decays within the framework of NRQCD. The numerical analysis of the decay width, the invariant mass and angular differential distributions, as well as the theoretical uncertainty are given in Sec.III. Finally Sec.IV gives a summary and conclusions.

\section{Calculation technology}

By the interaction of Higgs with heavy quarks in SM, the Feynman diagrams for the production of $\Xi_{cc}$, $\Xi_{bc}$ and $\Xi_{bb}$ are analogous with different heavy quark accordingly. For convenience, we use $\Xi_{QQ'}$ instead, $Q$ and $Q'$ are both represent the heavy quark $c$ or $b$, corresponding to the production of $\Xi_{cc}$, $\Xi_{bc}$ and $\Xi_{bb}$. 
\begin{figure}[htb]
  \centering
  \subfigure[]{
    \includegraphics[scale=0.27]{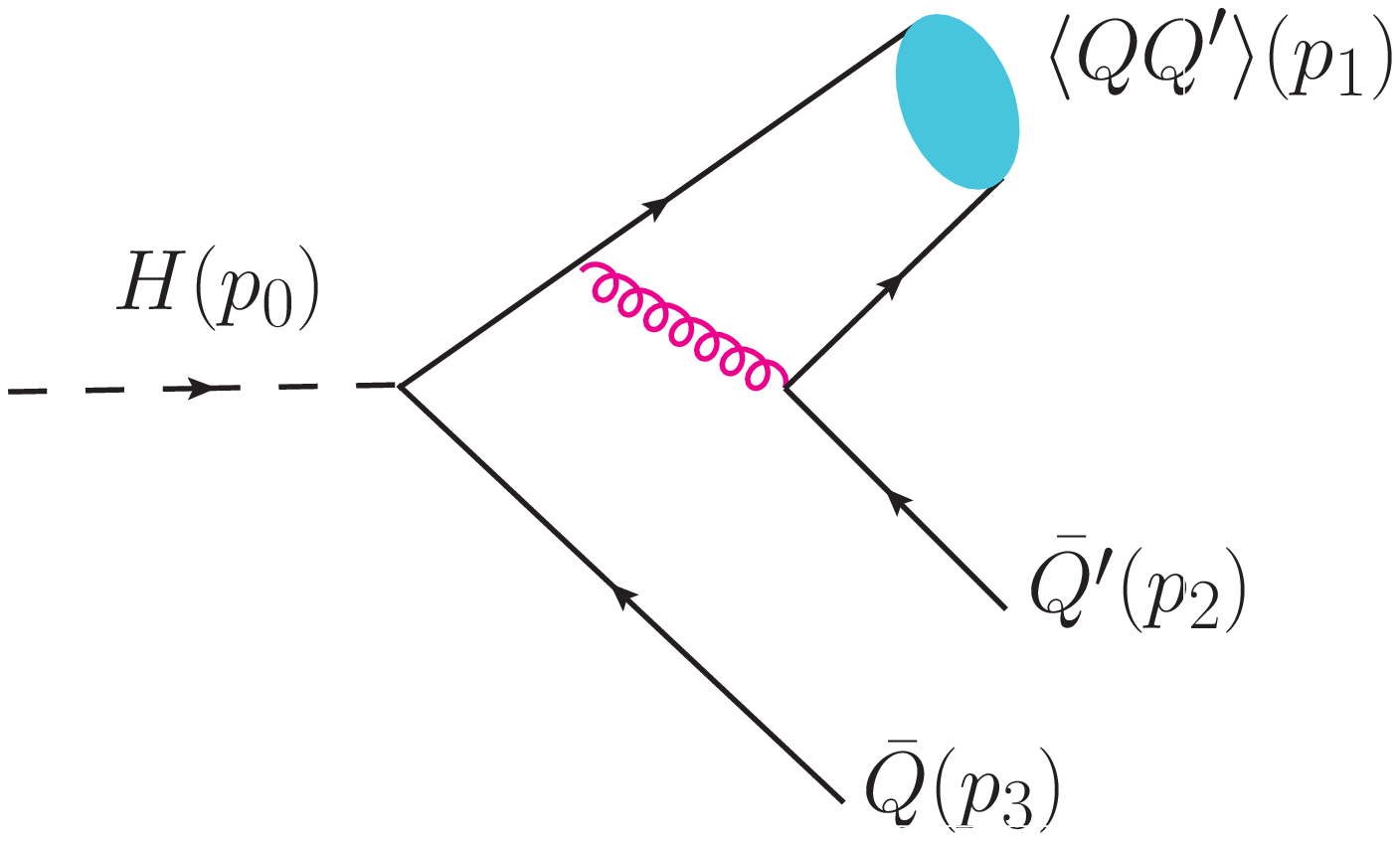}}
  \hspace{0.00in}
  \subfigure[]{
    \includegraphics[scale=0.27]{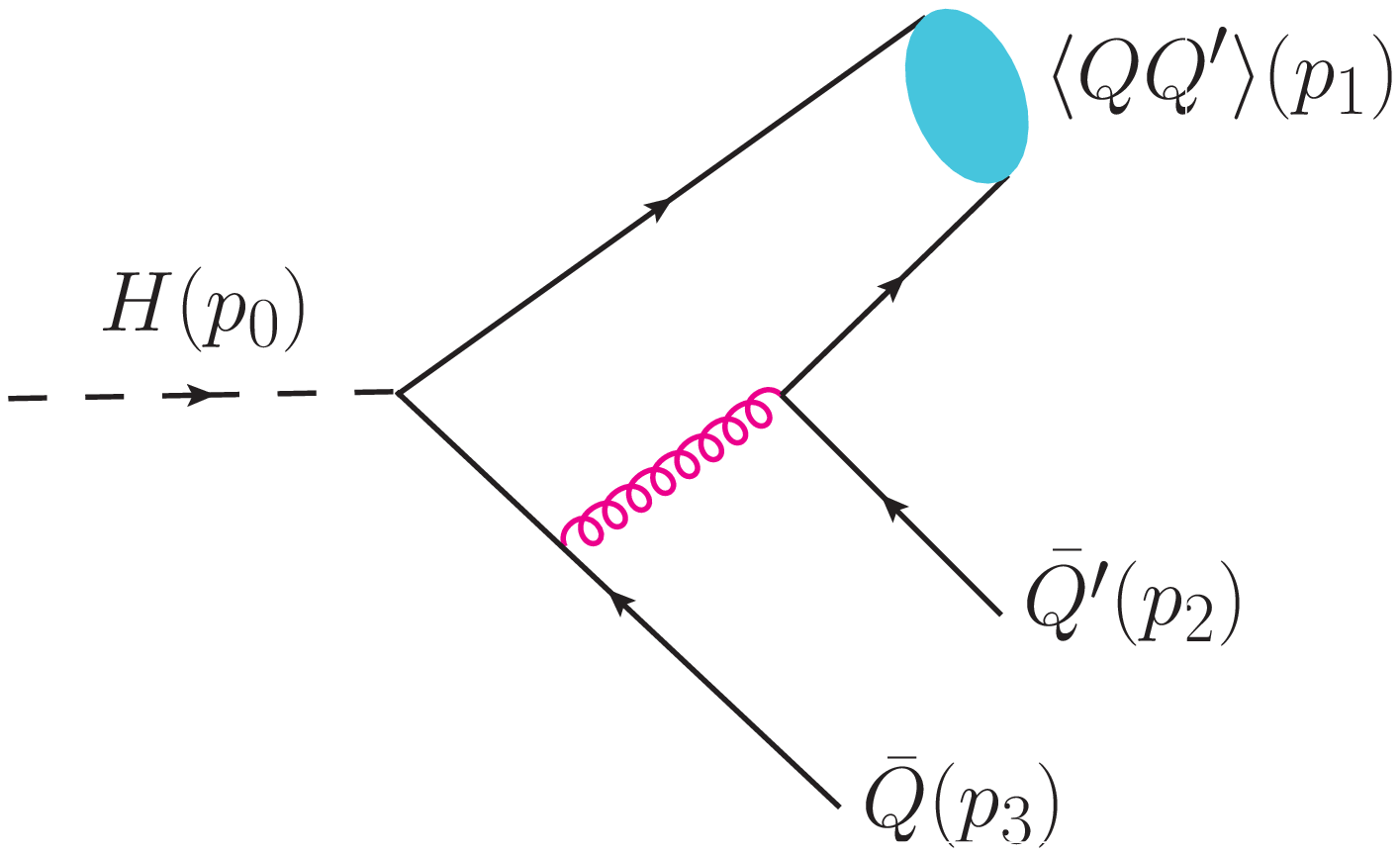}}
  \hspace{0.00in}
  \subfigure[]{
    \includegraphics[scale=0.33]{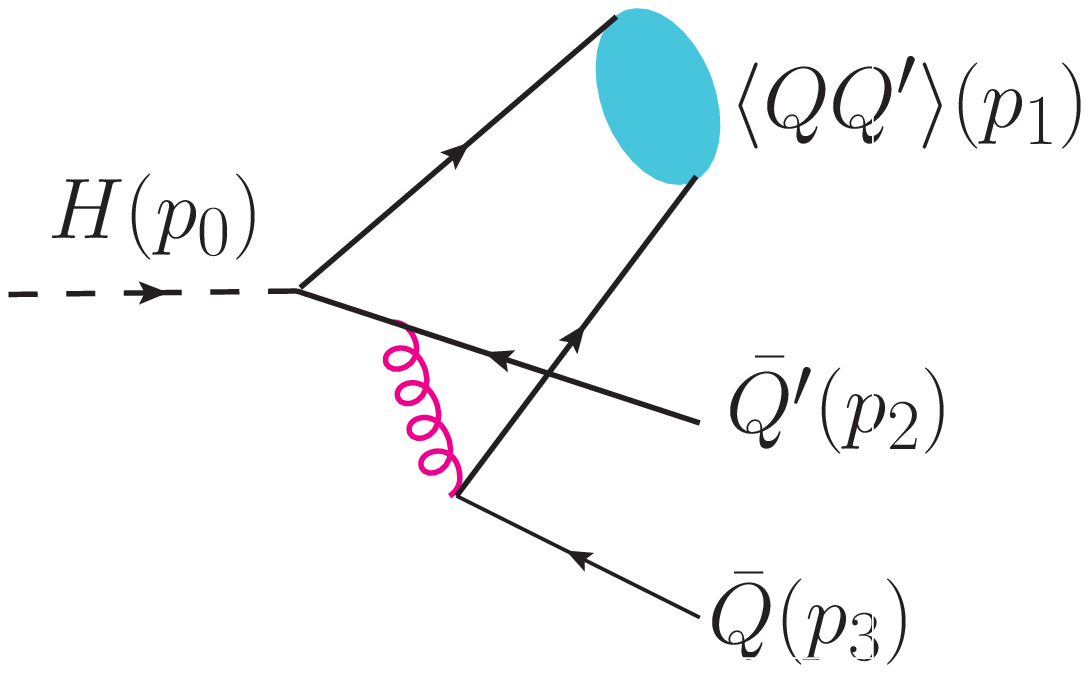}}
  \hspace{0.00in}
  \subfigure[]{
    \includegraphics[scale=0.31]{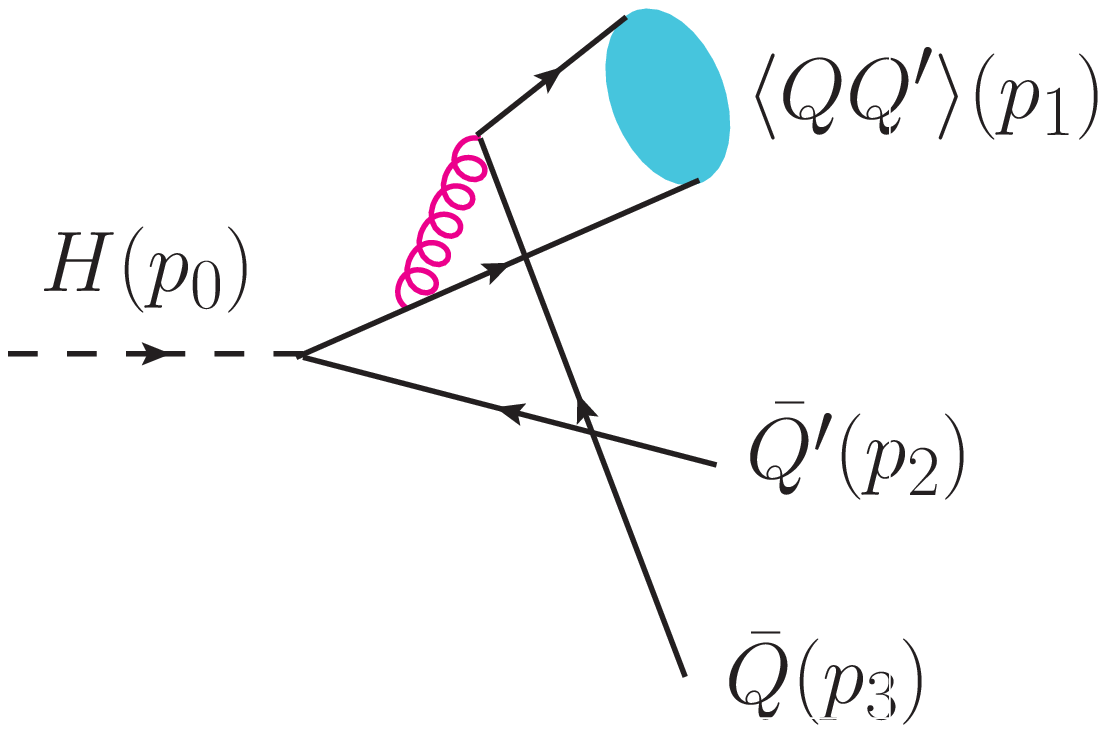}}
  \caption{Feynman diagrams for the process $H(p_0) \rightarrow \langle QQ'\rangle[n](p_1)+ \bar {Q^{\prime}} (p_2) + \bar{Q} (p_3)$, where $Q$ and $Q^{\prime}$ denote as the heavy $c$ or $b$ quark, $[n]$ is the spin and color quantum number of the diquark state, i.e., $\langle cc\rangle[^{1}P_{1}]_{\mathbf{\bar 3}}$, $\langle cc\rangle[^{3}P_{J}]_{\mathbf{6}}$, 
$\langle bc\rangle[^{1}P_{1}]_{\mathbf{\bar 3}/ \mathbf{6}}$, $\langle bc\rangle[^{3}P_{J}]_{\mathbf{\bar 3}/ \mathbf{6}}$, 
$\langle bb \rangle[^{1}P_{1}]_{\mathbf{\bar 3}}$ and $\langle bb\rangle[^{3}P_{J}]_{\mathbf{6}}$, with $J=0, 1, 2$ for the $P$-wave diquark state.}
  \label{diagram1}
\end{figure}

Based on the NRQCD theory~\cite{Bodwin:1994jh,Petrelli:1997ge}, the production of doubly heavy baryon $\Xi_{QQ'}$ in Higgs decays can be factorized into two parts, the perturbative short-distance coefficient for the production of intermediate diquark state $\langle QQ'\rangle[n]$ and the non-perturbative long-distance matrix elements, which represents the transition probability from the intermediate diquark state $\langle QQ' \rangle[n]$ to the corresponding doubly heavy baryons $\Xi_{QQ'}$. Specifically, the decay width can be factorized as
\begin{eqnarray}
\Gamma (H(p_0) &\rightarrow& \Xi_{QQ'}(p_1)+ \bar {Q'}(p_2) + \bar {Q}(p_3)) \nonumber\\
&=&\sum_{n} \hat{\Gamma}(H(p_0) \rightarrow \langle QQ'\rangle[n](p_1) + \bar {Q'}(p_2) + \bar {Q}(p_3)) \langle\mathcal O^{H}[n]\rangle,
\label{factorization}
\end{eqnarray}
where $[n]$ refers to the spin and color quantum number of the intermediate diquark state $\langle QQ^{\prime}\rangle$. To be specific, the spin quantum number of the orbital excited $P$-wave diquark state can be $[^1P_1]$ or $[^3P_J]$ with $J=0,1,2$. The color quantum number can be color-antitriplet $\mathbf{\bar 3}$ or color-sextuplet $\mathbf{6}$ for the decomposition of $SU_C(3)$ color group $\mathbf{3}\bigotimes \mathbf{3}=\mathbf{\bar 3} \bigoplus \mathbf{6}$. Thus for the production of excited $P$-wave doubly heavy baryon $\Xi_{bc}$, there are eight color and spin states for the diquark, such as $\langle bc\rangle[^{1}P_{1}]_{\mathbf{\bar 3}}$, $\langle bc\rangle[^{1}P_{1}]_{\mathbf{6}}$, $\langle bc\rangle[^{3}P_{J}]_{\mathbf{\bar 3}}$ and $\langle bc\rangle[^{3}P_{J}]_{\mathbf{6}}$. While for the production of $\Xi_{cc}$ ($\Xi_{bb}$) baryons, the intermediate $P$-wave diquark states can be $\langle cc\rangle[^{1}P_{1}]_{\mathbf{\bar 3}}$ and $\langle cc\rangle[^{3}P_{J}]_{\mathbf{6}}$ ($\langle bb \rangle[^{1}P_{1}]_{\mathbf{\bar 3}}$ and $\langle bb\rangle[^{3}P_{J}]_{\mathbf{6}}$) for the symmetry of identical particles in the diquark state.
$\langle\mathcal O^{H}[n]\rangle$ is the non-perturbative long-distance matrix element and represents the hadronic process from diquark state $\langle QQ'\rangle[n]$ to the produced $\Xi_{QQ^{\prime}}$. 
Assuming the potential of the diquark state in color-antitriplet are hydrogen-like, the transition probability $h_{\mathbf{\bar{3}}}$ can be approximatively related to the first derivative of the Schr\"{o}dinger wave function at the origin $|\Psi'_{QQ'}(0)|$ for all the considered $P$-wave states, which can be naturally connected to the first derivative of the radial wave function at the origin,
\begin{eqnarray}
\langle\mathcal O^{H}[P]\rangle=h_{\mathbf{\bar{3}}}=|\Psi'_{QQ'}(0)|^2=\frac{3}{4\pi}|R'_{QQ'}(0)|^2.
\end{eqnarray}
Nevertheless, there is no such a unambiguous relationship between the transition probability of diquark in color-sextuplet $h_{\mathbf{6}}$ and the wave function. In Refs.~\cite{Ma:2003zk,Chang:2006eu,Chang:2006xp,Jiang:2012jt,Niu:2019xuq,Niu:2018ycb}, it has been indicated that these two parameters $h_{\mathbf{\bar{3}}}$ and $h_{\mathbf{6}}$ are at the
same order of the relative velocity, $v_r$, between the two heavy
quarks in the baryon rest frame. So to reduce the selection of parameters, we assume $h_{\mathbf{6}}=h_\mathbf{\bar{3}}$ in the numerical calculation.

Next we would focus on the perturbative short-distance coefficients, the decay width $ \hat{\Gamma}(H(p_0) \rightarrow \langle QQ'\rangle[n](p_1) + \bar {Q'}(p_2) + \bar {Q}(p_3))$ in Eq.~(\ref{factorization}). Typical Feynman diagrams for the process $H (p_0) \rightarrow \langle QQ'\rangle[n](p_1)+ \bar{Q^{\prime}}(p_2) + \bar{Q}(p_3)$ are presented in Fig.~\ref{diagram1}, where $\langle QQ'\rangle[n]$ can be $\langle cc\rangle[^{1}P_{1}]_{\mathbf{\bar 3}}$, $\langle cc\rangle[^{3}P_{J}]_{\mathbf{6}}$, 
$\langle bc\rangle[^{1}P_{1}]_{\mathbf{\bar 3}/ \mathbf{6}}$, $\langle bc\rangle[^{3}P_{J}]_{\mathbf{\bar 3}/ \mathbf{6}}$, 
$\langle bb \rangle[^{1}P_{1}]_{\mathbf{\bar 3}}$ and $\langle bb\rangle[^{3}P_{J}]_{\mathbf{6}}$, with $J=0, 1, 2$ for the $P$-wave diquark state.
The decay width can be written as
\begin{eqnarray}
\hat{\Gamma}(H^0 \rightarrow \langle QQ'\rangle[n] + \bar {Q'} + \bar {Q})= \int \frac{1}{2m_H} \sum |\mathcal{M}[n]|^2 d\Phi_3,
\label{width}
\end{eqnarray}
where $m_H$ is the mass of Higgs, $\sum$ means to sum over the spin and color of the final-state particles. $\mathcal{M}[n]$ is the amplitude for the production of $P$-wave diquark state with spin and color quantum number $[n]$.
After 3-body phase space $d\Phi_3$ integral, Eq.~(\ref{width}) can be written as
\begin{eqnarray}\label{dw}
d \hat{\Gamma} = \frac{1}{256 \pi^{3} m_{H}^{3}} \sum |\mathcal{M}[n]|^2 ds_{12} ds_{23},
\end{eqnarray}
in which the invariant mass $\rm s_{ij}=\it (p_{i}+p_{j})^{\rm{2}}$ with $i,j=1,2,3$. With the help of Eq.~(\ref{dw}), not only can the decay width of all considered diquark states be obtained, but also the corresponding differential distributions can be obtained to analyze the kinematic characteristics. 

For the production of diquark state $\langle QQ'\rangle[n]$ within the NRQCD framework, a more accustomed action needs to be applied to the amplitude $\mathcal{M}[n]$, that is the charge parity $C=-i\gamma^2\gamma^0$. After the action of charge parity, $\mathcal M[n]$ can be related to the amplitude of the process $H(p_0) \rightarrow (Q\bar {Q^{\prime}})[n](p_1) + Q^{\prime}(p_2) + \bar{Q}(p_3)$ with an additional factor $(-1)^{m+1}$, where $m$ stands for the number of vector vertices in the $Q'$ fermion line which is the opposite of that for the diquark production, and here $m$ = 1. The proof of the charge parity acting on the fermion line $Q'$ has been shown in Refs.~\cite{Jiang:2012jt,Zheng:2015ixa} in detail. 
According to Fig.~\ref{diagram1}, the hard amplitude $\mathcal M[n]$ of the $P$-wave diquark states can be obtained by taking the first derivative of the relative momentum $p$ in the amplitude of the $S$-wave diquark state. Clearly, they can be written as:
\begin{widetext}
\begin{eqnarray}\label{m1p11}
\mathcal{M}_{a}[^1P_1]&=& \kappa_Q \left. \varepsilon^l_{\alpha}(p_1)\frac{d}{dp_{\alpha}}\left[ \bar{u}_i(p_2) \gamma_{\mu}  \frac{\Pi_{p_1}[^1 S_0]}{(p_2+p_{12})^2} \gamma_{\mu} \frac{\slashed{p}_{1}+\slashed{p}_{2}+m_Q}{(p_1+p_2)^2-m_{Q}^{2}}  v_j(-p_3)\right]\right|_{p=0}, \\
\mathcal{M}_{b}[^1P_1]&=& \kappa_Q  \left. \varepsilon^l_{\alpha}(p_1)\frac{d}{dp_{\alpha}}\left[ \bar{u}_i(p_2) \gamma_{\mu}  \frac{\Pi_{p_1}[^1 S_0]}{(p_2+p_{12})^2}  \frac{-\slashed{p}_{12}-\slashed{p}_{2}-\slashed{p}_{3}+m_Q}{(p_{12}+p_2+p_3)^2-m_{Q}^{2}} \gamma_{\mu} v_j(-p_3)\right]\right|_{p=0},  \\
\mathcal{M}_{c}[^1P_1]&=& \kappa_{Q^{\prime}} \left. \varepsilon^l_{\alpha}(p_1)\frac{d}{dp_{\alpha}}\left[ \bar{u}_i(p_2) \gamma_{\mu}  \frac{\slashed{p}_{11}+\slashed{p}_{2}+\slashed{p}_{3}+m_{Q^{\prime}}}{(p_{11}+p_2+p_3)^2-m_{Q^{\prime}}^{2}}  \frac{\Pi_{p_1}[^1 S_0]}{(p_3+p_{11})^2}  \gamma_{\mu} v_j(-p_3)\right]\right|_{p=0}, \\
\mathcal{M}_{d}[^1P_1]&=& \kappa_{Q^{\prime}} \left. \varepsilon^l_{\alpha}(p_1)\frac{d}{dp_{\alpha}}\left[ \bar{u}_i(p_2) \frac{-\slashed{p}_{1}-\slashed{p}_{3}+m_{Q^{\prime}}}{(p_{1}+p_3)^2-m_{Q^{\prime}}^{2}}
\gamma_{\mu} \frac{\Pi_{p_1}[^1 S_0]}{(p_3+p_{11})^2} \gamma_{\mu} v_j(-p_3)\right]\right|_{p=0},\label{m1p14}
\end{eqnarray}
\end{widetext}
and
\begin{widetext}
\begin{eqnarray}\label{m3pj1}
\mathcal{M}_{a}[^3P_J]&=& \kappa_Q \left. \varepsilon^J_{\alpha \beta}(p_1)\frac{d}{dp_{\alpha}}\left[ \bar{u}_i(p_2)\gamma_{\mu}  \frac{\Pi_{p_1}^{\beta}[^3 S_1]}{(p_2+p_{12})^2} \gamma_{\mu} \frac{\slashed{p}_{1}+\slashed{p}_{2}+m_Q}{(p_1+p_2)^2-m_{Q}^{2}}  v_j(-p_3)\right]\right|_{p=0}, \\
\mathcal{M}_{b}[^3P_J]&=& \kappa_Q \left. \varepsilon^J_{\alpha \beta}(p_1)\frac{d}{dp_{\alpha}}\left[ \bar{u}_i(p_2) \gamma_{\mu}  \frac{\Pi_{p_1}^{\beta}[^3 S_1]}{(p_2+p_{12})^2}   \frac{-\slashed{p}_{12}-\slashed{p}_{2}-\slashed{p}_{3}+m_Q}{(p_{12}+p_2+p_3)^2-m_{Q}^{2}} \gamma_{\mu} v_j(-p_3)\right]\right|_{p=0},  \\
\mathcal{M}_{c}[^3P_J]&=& \kappa_{Q^{\prime}} \left. \varepsilon^J_{\alpha \beta}(p_1)\frac{d}{dp_{\alpha}}\left[ \bar{u}_i(p_2) \gamma_{\mu}  \frac{\slashed{p}_{11}+\slashed{p}_{2}+\slashed{p}_{3}+m_{Q^{\prime}}}{(p_{11}+p_2+p_3)^2-m_{Q^{\prime}}^{2}} \frac{\Pi_{p_1}^{\beta}[^3 S_1]}{(p_3+p_{11})^2} \gamma_{\mu} v_j(-p_3)\right]\right|_{p=0}, \\
\mathcal{M}_{d}[^3P_J]&=& \kappa_{Q^{\prime}} \left. \varepsilon^J_{\alpha \beta}(p_1)\frac{d}{dp_{\alpha}}\left[\bar{u}_i(p_2) \frac{-\slashed{p}_{1}-\slashed{p}_{3}+m_{Q^{\prime}}}{(p_{1}+p_3)^2-m_{Q^{\prime}}^{2}}
\gamma_{\mu} \frac{\Pi_{p_1}^{\beta}[^3 S_1]}{(p_3+p_{11})^2} \gamma_{\mu} v_j(-p_3)\right]\right|_{p=0},\label{m3pj4}
\end{eqnarray}
\end{widetext}
with
\begin{widetext}
\begin{eqnarray}\label{overallf}
\kappa_{Q/Q^{\prime}}=\mathcal{C}  \frac{e g_s^2 m_{Q/Q^{\prime}}}{2 m_W \sin\theta_W},
\end{eqnarray}
\end{widetext}
in which $\theta_W$ refers to the Weinberg angle. The color factor $\mathcal{C}$ can be described as
\begin{eqnarray}
\mathcal{C}_{ij,k}=\mathcal{N} \times \sum_{a,m,n} (T^a)_{mi} (T^a)_{nj} \times G_{mnk},
\end{eqnarray}
where the normalization constant $\mathcal{N}=\sqrt{1/2}$; $i, j, m, n= 1, 2, 3$ are the color indices of the outgoing $\bar{Q}$, $\bar{Q'}$, $Q$ and $Q^{\prime}$, respectively; $a=1, \ldots, 8$ and $k$ denote the color indices of the gluon and the diquark state $\langle QQ^{\prime}\rangle$; $G_{mnk}$ is the antisymmetric function $\varepsilon_{mnk}$ for color-antitriplet state, or the symmetric function $f_{mnk}$ for color-sextuplet state. For the square of color factor $\mathcal{C}^{2}_{ij,k}$, $G_{mnk}$ satisfy
\begin{eqnarray}
\varepsilon_{mnk} \varepsilon_{m^{\prime}n^{\prime}k}=\delta_{mm^{\prime}}\delta_{nn^{\prime}}-\delta_{mn^{\prime}}\delta_{nm^{\prime}},
\nonumber\\f_{mnk} f_{m^{\prime}n^{\prime}k}=\delta_{mm^{\prime}}\delta_{nn^{\prime}}+\delta_{mn^{\prime}}\delta_{nm^{\prime}},
\end{eqnarray}
resulting $\mathcal{C}^{2}$ equals $\frac{4}{3}$ for the production of color-antitriplet diquark state, and $\frac{2}{3}$ for the color-sextuplet diquark state.

To ensure the gauge invariance, the mass of diquark $M_{QQ'}$ is taken to be $m_{Q}+ m_{Q'}$. $p_{11}$ and $p_{12}$ are the momenta of the two constituent quarks in the diquark state, and explicitly they can be $p_{11}=\frac{m_{Q}}{M_{QQ^{\prime}}}p_{1}+p$ and $\it p_{\rm{12}}=\frac{m_{Q'}}{M_{QQ'}}p_{\rm 1}-p$,
where $p$ is the relative momentum between these two constituent quarks. According to the non-relativistic approximation, the relative momentum $p$ is small enough to be neglected in the final amplitude. $\varepsilon^l_{\alpha}$ and $\varepsilon^J_{\alpha\beta}$ are the polarization vector of the $^1P_1$ diquark state and polarization tensor of the $^3P_J$ diquark state with $J=0, 1, 2$, respectively, and they need to be summed by polarization sum formula to select the proper total angular momentum. Therefore, the sum formula of polarization vector and polarization tensor should be taken as the following forms~\cite{Petrelli:1997ge}
\begin{eqnarray}
	\sum_{l_{z}} \varepsilon^l_{\alpha} \varepsilon_{\alpha^{\prime}}^{l*} &=&\Pi_{\alpha \alpha^{\prime}},\\
	\varepsilon_{\alpha \beta}^{0} \varepsilon_{\alpha^{\prime} \beta^{\prime}}^{0*} &=&\frac{1}{3} \Pi_{\alpha \beta} \Pi_{\alpha^{\prime} \beta^{\prime}}, \label{psum3p0}\\
	\sum_{J_{z}} \varepsilon_{\alpha \beta}^{1} \varepsilon_{\alpha^{\prime} \beta^{\prime}}^{1 *} &=&\frac{1}{2}\left(\Pi_{\alpha \alpha^{\prime}} \Pi_{\beta \beta^{\prime}}-\Pi_{\alpha \beta^{\prime}} \Pi_{\alpha^{\prime} \beta}\right), \label{psum3p1}\\
	\sum_{J_{z}} \varepsilon_{\alpha \beta}^{2} \varepsilon_{\alpha^{\prime} \beta^{\prime}}^{2 *} &=&\frac{1}{2}\left(\Pi_{\alpha \alpha^{\prime}} \Pi_{\beta \beta^{\prime}}+\Pi_{\alpha \beta^{\prime}} \Pi_{\alpha^{\prime} \beta}\right)-\frac{1}{3} \Pi_{\alpha \beta} \Pi_{\alpha^{\prime} \beta^{\prime}},\label{psum3p2}
\end{eqnarray}
with the definition 
\begin{eqnarray}
	\Pi_{\alpha \beta}=-g_{\alpha \beta}+\frac{p_{1 \alpha} p_{1 \beta}}{M_{QQ'}^{2}}.
\end{eqnarray}

The projector $\Pi_{p_1}[^1 S_0]$ in Eqs. (\ref{m1p11}-\ref{m1p14}), $\Pi_{p_1}^{\beta}[^3 S_1]$ in Eqs. (\ref{m3pj1}-\ref{m3pj4}) and its first derivative of the relative momentum $p$ are
\begin{widetext}
\begin{eqnarray}
\Pi_{p_1}[^1 S_0] &=& \frac{-\sqrt{M_{QQ^{\prime}}}}{4 m_{Q} m_{Q^{\prime}}}\left(\slashed{p}_{12}-m_{Q^{\prime}}\right) \gamma^{5}\left(\slashed{p}_{11}+m_{Q}\right),\\
\Pi_{p_1}^{\beta}[^3 S_1] &=& \frac{-\sqrt{M_{QQ^{\prime}}}}{4 m_{Q} m_{Q^{\prime}}}\left(\slashed{p}_{12}-m_{Q^{\prime}}\right) \gamma^{\beta} \left(\slashed{p}_{11}+m_{Q}\right),\\
\left. \frac{d}{dp_{\alpha}}\Pi_{p_1}[^1 S_0] \right|_{p=0}&=& \frac{\sqrt{M_{QQ^{\prime}}}}{4 m_{Q} m_{Q^{\prime}}}\gamma^{\alpha} \gamma^{5} \left(\slashed{p}_{1}+m_{Q}-m_{Q^{\prime}}\right), \\
\left. \frac{d}{dp_{\alpha}}\Pi_{p_1}^{\beta}[^3 S_1] \right|_{p=0}&=& \frac{\sqrt{M_{QQ^{\prime}}}}{4 m_{Q} m_{Q^{\prime}}} \left[ \gamma^{\alpha} \gamma^{\beta} \left(\slashed{p}_{1}+m_{Q}-m_{Q^{\prime}}\right) -2 g^{\alpha \beta} \left(\slashed{p}_{12} - m_{Q^{\prime}} \right) \right].
\end{eqnarray}
\end{widetext}

\section{Numerical results}

In the numerical calculation, the mass of Higgs $m_H=\rm 125.18~{GeV}$ and the total decay width of Higgs is considered as $\Gamma_H=4.2$~MeV~\cite{Heinemeyer:2013tqa} to estimate the branching ratio and corresponding events for the production of $\Xi_{QQ'}$. The masses of heavy quarks are taken as $m_c=1.8~\rm{GeV}$ and $m_b=\rm 5.1~{GeV}$ to build the masses of doubly heavy baryons, i.e., $M_{\Xi_{cc}}=3.6~{\rm GeV}$, $M_{\Xi_{bc}}=6.9~{\rm GeV}$ and $M_{\Xi_{bb}}=10.2~{\rm GeV}$. The first derivative of the radial wave function $|R'_{cc}(0)|=~{\rm 0.102~GeV}^{5/2}$, $|R'_{bc}(0)|=0.200~{\rm GeV}^{5/2}$, and $|R'_{bb}(0)|=0.479~{\rm GeV}^{5/2}$, which are the same as that in Ref.~\cite{Kiselev:2002iy} and obtained under the $K^2O$ potential motivated by QCD with a three-loop function.
The running strong coupling is adopted to be $\alpha_s(\mu_r=2m_c) = 0.242$ for the production of $\Xi_{cc}$ and $\Xi_{bc}$, and $\alpha_s(\mu_r=2m_b) = 0.180$ for $\Xi_{bb}$ production.
The $G_{F}$ scheme is adopted to calculate the electroweak coupling constant with

\begin{eqnarray}
&&\it G_{F} \rm =1.1663787 \times 10^{-5}~GeV^{-2}, ~\cos\theta_W=\frac{\it m_W}{\it m_Z},\nonumber\\
&&m_Z=91.1876~\rm{GeV},~~~~~~~~~~\it{m_W}=\rm 80.385~{GeV}.
\end{eqnarray}

The programs FeynArts 3.9~\cite{Hahn:2000kx} and FeyCalc 9.3~\cite{Shtabovenko:2020gxv} are used to generate the amplitudes and do the algebraic and numerical calculations.

\subsection{Decay widths}

Based on the parameters mentioned before, the decay widths of all considered $P$-wave diquark states, i.e., $\langle cc\rangle[^{1}P_{1}]_{\mathbf{\bar 3}}$, $\langle cc\rangle[^{3}P_{J}]_{\mathbf{6}}$, $\langle bc\rangle[^{1}P_{1}]_{\mathbf{\bar 3/6}}$, $\langle bc\rangle[^{3}P_{J}]_{\mathbf{\bar 3/6}}$, $\langle bb \rangle[^{1}P_{1}]_{\mathbf{\bar 3}}$ and $\langle bb\rangle[^{3}P_{J}]_{\mathbf{6}}$ with $J=0,1,2$,  can be calculated for the production of excited $\Xi_{cc}$, $\Xi_{bc}$ and $\Xi_{bb}$, which are listed in Tables.~\ref{ccwidth}-\ref{bbwidth}, respectively. The contributions of the $S$-wave diquark states, i.e., $\langle cc\rangle[^{1}S_{0}]_{\mathbf{6}}$, $\langle cc\rangle[^{3}S_{1}]_{\mathbf{\bar 3}}$, $\langle bc\rangle[^{1}S_{0}]_{\mathbf{\bar 3/6}}$, $\langle bc\rangle[^{3}S_{1}]_{\mathbf{\bar 3/6}}$, $\langle bb \rangle[^{1}S_{0}]_{\mathbf{6}}$ and $\langle bb\rangle[^{3}S_{1}]_{\mathbf{\bar 3}}$ are also added in the corresponding Table.~\ref{ccwidth},~\ref{bcwidth}, or~\ref{bbwidth} for comprehensive comparison and analysis.

The produced $\Xi_{QQ'}$ events can be roughly estimated by formula $N_{\Xi_{QQ'}[n]}=N_{H}\rm{Br}_{\Xi_{QQ'}[n]}$ at the future colliders, the so-called ``Higgs factories''. Here the branching ratio of Higgs decay channel is defined as
\begin{eqnarray}
\rm {Br}_{\Xi_{QQ'}[n]}=\frac{\Gamma_{H \rightarrow \Xi_{QQ'}[n] + \bar{Q'} + \bar{Q}}}{\Gamma_H},
\end{eqnarray}
and $N_{H}$ stands for the total Higgs events produced at the future colliders. 
Specifically, HL-LHC running at center-of-mass collision energy $\sqrt{s}=14$~TeV with the integrated luminosity $3~ab^{-1}$ would produce about $1.65\times10^{8}$ Higgs events per year \cite{LHCHIGGS}. 
For the upgraded HE-LHC running at $\sqrt{s}=33$ TeV, the cross-section would be as high as 200 pb, resulting in the Higgs events produced per year up to $6.0\times10^{8}$. 
At the $e^{+}e^{-}$ colliders, more than one million Higgs events would be produced at CEPC with the center-of-mass energy $\sqrt{s}=240$~GeV and the integrated luminosity $0.8~ab^{-1}$ in 7 years~\cite{CEPCStudyGroup:2018rmc}, and almost the same magnitude of Higgs bosons would be produced at the ILC~\cite{Simon:2012ik}.
The branching ratios (Br for short in  the following tables) and the estimated $\Xi_{QQ'}$ events produced at HL-LHC and CEPC/ILC are showed in Table.~\ref{ccwidth},~\ref{bcwidth}, and~\ref{bbwidth}, respectively. $S$-wave ($P$-wave) in Tables.~\ref{ccwidth}-\ref{bbwidth} denotes the contributions for the sum of states whose orbital quantum number is $[S] ([P])$. Suppose that all the highly excited states can decay into the ground state 100\%, the total decay width could be obtained by summing up all the $S$-wave and $P$-wave contributions, which are also added in Tables.~\ref{ccwidth}-\ref{bbwidth}.

\begin{table}[htb]
\begin{center}
\caption{Decay widths ($\times 10^{-9}$ GeV), branching ratios (Br), and produced events at HL-LHC and CEPC/ILC for $P$-wave and $S$-wave $\Xi_{cc}$.}
	\begin{tabular}{|c||c||c|c|c|}
		\hline
		$[n]$&Decay width&Br $(\times 10^{-6})$&HL-LHC events &CEPC/ILC events\\
		\hline\hline
		$[^1S_0]_{\mathbf{6}}$     & 34.68 & $8.26 \times 10^{0}$ & $1.36 \times 10^{3}$ & $8.26 \times 10^{0}$  \\
		\hline
		$[^3S_1]_{\overline{\mathbf{3}}}$     & 65.44 & $1.56 \times 10^{1}$ & $2.57 \times 10^{3}$ & $1.56 \times 10^{1}$  \\
		\hline
		$[^1P_1]_{\overline{\mathbf{3}}}$     & 1.08 & $2.56 \times 10^{-1}$ & $4.23 \times 10^{1}$ & $2.56 \times 10^{-1}$  \\
		\hline
		$[^3P_0]_{\mathbf{6}}$     & 0.83 & $1.97 \times 10^{-1}$ & $3.25 \times 10^{1}$ & $1.97 \times 10^{-1}$  \\
		\hline
		$[^3P_1]_{\mathbf{6}}$     & 0.86 & $2.04 \times 10^{-1}$ & $3.36 \times 10^{1}$ & $2.04 \times 10^{-1}$  \\
		\hline
		$[^3P_2]_{\mathbf{6}}$     & 0.29 & $7.00 \times 10^{-2}$ & $1.16 \times 10^{1}$ & $7.00 \times 10^{-2}$  \\
		\hline
		$S$-wave    & 100.12 & $2.38 \times 10^{1}$ & $3.93 \times 10^{3}$ & $2.38 \times 10^{1}$  \\
		\hline
		$P$-wave     & 3.05 & $7.27 \times 10^{-1}$ & $1.20 \times 10^{2}$ & $7.27 \times 10^{-1}$  \\
		\hline
		Total    & 103.18 & $2.46 \times 10^{1}$ & $4.05 \times 10^{3}$ & $2.46 \times 10^{1}$  \\
		\hline
	\end{tabular}
	\label{ccwidth}
	\end{center}
\end{table}

\begin{table}[htb]
\begin{center}
\caption{Decay widths ($\times 10^{-9}$ GeV), branching ratios (Br), and produced events at HL-LHC and CEPC/ILC for $P$-wave and $S$-wave $\Xi_{bc}$.}
	\begin{tabular}{|c||c||c|c|c|}
		\hline
		~~$[n]$~~&~Decay width~&~Br $(\times 10^{-6})$~&~HL-LHC events ~&~CEPC/ILC events~\\
		\hline\hline
		$[^1S_0]_{\overline{\mathbf{3}}}$     & 457.12 & $1.09 \times 10^{2}$ & $1.80 \times 10^{4}$ & $1.09 \times 10^{2}$  \\
		\hline
		$[^1S_0]_{\mathbf{6}}$     & 228.56 & $5.44 \times 10^{1}$ & $8.98 \times 10^{3}$ & $5.44 \times 10^{1}$  \\
		\hline
		$[^3S_1]_{\overline{\mathbf{3}}}$     & 586.79 & $1.40 \times 10^{2}$ & $2.31 \times 10^{4}$ & $1.40 \times 10^{2}$  \\
		\hline
		$[^3S_1]_{\mathbf{6}}$     & 293.63 & $6.99 \times 10^{1}$ & $1.15 \times 10^{4}$ & $6.99 \times 10^{1}$ \\
		\hline
		$[^1P_1]_{\overline{\mathbf{3}}}$     & 6.60 & $1.57 \times 10^{0}$ & $2.59 \times 10^{2}$ & $1.57 \times 10^{0}$  \\
		\hline
		$[^1P_1]_{\mathbf{6}}$     & 3.30 & $7.86 \times 10^{-1}$ & $1.30 \times 10^{2}$ & $7.86 \times 10^{-1}$  \\
		\hline
		$[^3P_0]_{\overline{\mathbf{3}}}$     & 4.07 & $9.69 \times 10^{-1}$ & $1.60 \times 10^{2}$ & $9.69 \times 10^{-1}$  \\
		\hline
		$[^3P_0]_{\mathbf{6}}$     & 2.03 & $4.84 \times 10^{-1}$ & $7.99 \times 10^{1}$ & $4.84 \times 10^{-1}$  \\
		\hline
		$[^3P_1]_{\overline{\mathbf{3}}}$     & 16.07 & $3.83 \times 10^{0}$ & $6.31 \times 10^{2}$ & $3.83 \times 10^{0}$ \\
		\hline
		$[^3P_1]_{\mathbf{6}}$     & 8.04 & $1.91 \times 10^{0}$ & $3.16 \times 10^{2}$ & $1.91 \times 10^{0}$  \\
		\hline
		$[^3P_2]_{\overline{\mathbf{3}}}$     & 6.95 & $1.65 \times 10^{0}$ & $2.73 \times 10^{2}$ & $1.65 \times 10^{0}$  \\
		\hline
		$[^3P_2]_{\mathbf{6}}$     & 3.47 & $8.27 \times 10^{-1}$ & $1.36 \times 10^{2}$ & $8.27 \times 10^{-1}$  \\
		\hline
		$S$-wave    & 1566.09 & $3.73 \times 10^{2}$ & $6.15 \times 10^{4}$ & $3.73 \times 10^{2}$  \\
		\hline
		$P$-wave     & 50.54 & $1.20 \times 10^{1}$ & $1.99 \times 10^{3}$ & $1.20 \times 10^{1}$  \\
		\hline
		Total    & 1616.63 & $3.85 \times 10^{2}$ & $6.35 \times 10^{4}$ & $3.85 \times 10^{2}$  \\
		\hline
	\end{tabular}
	\label{bcwidth}
	\end{center}
\end{table}

\begin{table}[htb]
\begin{center}
\caption{Decay widths ($\times 10^{-9}$ GeV), branching ratios (Br), and produced events at HL-LHC and CEPC/ILC for $P$-wave and $S$-wave $\Xi_{bb}$.}
\begin{tabular}{|c||c||c|c|c|}
\hline
~~$[n]$~~&~Decay width~&~Br $(\times 10^{-6})$~&~HL-LHC events ~&~CEPC/ILC events~\\
\hline\hline
$[^1S_0]_{\mathbf{6}}$     & 27.82 & $6.62 \times 10^{0}$ & $1.09 \times 10^{3}$ & $6.62 \times 10^{0}$  \\
\hline
$[^3S_1]_{\overline{\mathbf{3}}}$     & 40.63 & $9.67 \times 10^{0}$ & $1.60 \times 10^{3}$ & $9.67 \times 10^{0}$  \\
\hline
$[^1P_1]_{\overline{\mathbf{3}}}$     & 0.47 & $1.13 \times 10^{-1}$ & $1.86 \times 10^{1}$ & $1.13 \times 10^{-1}$  \\
\hline
$[^3P_0]_{\mathbf{6}}$     & 0.44 & $1.05 \times 10^{-1}$ & $1.72 \times 10^{1}$ & $1.05 \times 10^{-1}$  \\
\hline
$[^3P_1]_{\mathbf{6}}$     & 0.45 & $1.07 \times 10^{-1}$ & $1.76 \times 10^{1}$ & $1.07 \times 10^{-1}$  \\
\hline
$[^3P_2]_{\mathbf{6}}$     & 0.14 & $3.29 \times 10^{-2}$ & $5.43 \times 10^{0}$ & $3.29 \times 10^{-2}$  \\
\hline
$S$-wave    & 68.44 & $1.63 \times 10^{1}$ & $2.69 \times 10^{3}$ & $1.63 \times 10^{1}$  \\
\hline
$P$-wave     & 1.50 & $3.57 \times 10^{-1}$ & $5.89 \times 10^{1}$ & $3.57 \times 10^{-1}$  \\
\hline
Total    & 69.94 & $1.67 \times 10^{1}$ & $2.75 \times 10^{3}$ & $1.67 \times 10^{1}$  \\
\hline
\end{tabular}
\label{bbwidth}
\end{center}
\end{table}

Tables~\ref{ccwidth}-\ref{bbwidth} indicate that
\begin{itemize}
  \item For the production of $P$-wave $\Xi_{cc}$ and $\Xi_{bb}$, the contribution of the intermediate diquark state $[^1P_1]_{\overline{\mathbf{3}}}$ domains, while the contribution of $[^3P_1]_{\overline{\mathbf{3}}}$ state is the largest for the poduction of $P$-wave $\Xi_{bc}$. 
After further considering the contributions from $P$-wave states, the total decay widths through the corresponding process for the production of $\Xi_{cc}$, $\Xi_{bc}$, and $\Xi_{bb}$ would be
\begin{eqnarray}
&&\Gamma_{H \rightarrow \Xi_{cc}+ \bar {c} + \bar {c}}=103.18 \times 10^{-9}~\rm{GeV}, \nonumber\\
&&\Gamma_{H \rightarrow \Xi_{bc}+ \bar {c} + \bar {b}}=1616.63 \times 10^{-9}~\rm{GeV}, \nonumber\\
&&\Gamma_{H \rightarrow \Xi_{bb}+ \bar {b} + \bar {b}}=69.94 \times 10^{-9}~\rm{GeV}. \nonumber
\end{eqnarray}
  \item There are four intermediate states for the production of $\Xi_{cc}$ and $\Xi_{bb}$, that is, $[^{1}P_{1}]_{\mathbf{\bar 3}}$, $[^{3}P_{J}]_{\mathbf{6}}$, and eight for the production of $\Xi_{bc}$, namely, $\langle bc\rangle[^{1}P_{1}]_{\mathbf{\bar 3/6}}$, $\langle bc\rangle[^{3}P_{J}]_{\mathbf{\bar 3/6}}$. The ratios among these corresponding $P$-wave states are 
\begin{eqnarray}
&&\langle cc\rangle[^{1}P_{1}]_{\mathbf{\bar 3}}:\langle cc\rangle[^{3}P_{0}]_{\mathbf{6}}:\langle cc\rangle[^{3}P_{1}]_{\mathbf{6}}:\langle cc\rangle[^{3}P_{2}]_{\mathbf{6}} = 1:0.77:0.80:0.27, \nonumber\\
&&\langle bc\rangle[^{3}P_{1}]_{\mathbf{\bar 3}}:\langle bc\rangle[^{3}P_{1}]_{\mathbf{6}}:\langle bc\rangle[^{3}P_{0}]_{\mathbf{\bar 3}}:\langle bc\rangle[^{3}P_{0}]_{\mathbf{6}}:
\langle bc\rangle[^{3}P_{2}]_{\mathbf{\bar 3}}:\langle bc\rangle[^{3}P_{2}]_{\mathbf{6}}:\nonumber\\
&&\langle bc\rangle[^{1}P_{1}]_{\mathbf{\bar 3}}:\langle bc\rangle[^{1}P_{1}]_{\mathbf{6}}
= 1:0.5:0.25:0.13:0.43:0.22:0.41:0.21,\nonumber\\
&&\langle bb\rangle[^{1}P_{1}]_{\mathbf{\bar 3}}:\langle bb\rangle[^{3}P_{0}]_{\mathbf{6}}:\langle bb\rangle[^{3}P_{1}]_{\mathbf{6}}:\langle bb\rangle[^{3}P_{2}]_{\mathbf{6}} = 1:0.94:0.96:0.30.\nonumber
\end{eqnarray}
Each of these intermediate $P$-wave diquark states has a significant contribution and thus they all should be seriously considered for the production of excited $\Xi_{QQ^{\prime}}$.
  \item The branching ratio for the production of $P$-wave $\Xi_{cc}$ and $\Xi_{bb}$ baryons is about $10^{-7}$ which is almost two orders of magnitude lower than that of $S$-wave $\Xi_{cc}$ and $\Xi_{bb}$. While for the production of $P$-wave $\Xi_{bc}$ baryons, the Br is about $10^{-5}$ and it is one order of magnitude lower than that of $S$-wave $\Xi_{bc}$. 
  \item The estimated event number of produced doubly heavy baryons $\Xi_{bc}$ are the most and they are about one order magnitude larger than that of $\Xi_{cc}$ and $\Xi_{bb}$. And the events of $P$-wave $\Xi_{cc}$, $\Xi_{bc}$ and $\Xi_{bb}$ is respectively about 3.05\%, 3.23\% and 2.19\% of that for $S$-wave. Although the proportion of $P$-wave is small, the future colliders should be able to produce a sufficient number of $P$-wave doubly heavy baryon events along with so many events produced from $S$-wave. The events of $P$-wave can also be used as a supplement to $S$-wave.
  \item At HL-LHC, there are sizable doubly heavy baryon events produced per year, i.e., 0.41$\times10^4$ events of $\Xi_{cc}$, 6.35$\times10^4$ events of $\Xi_{bc}$ and 0.28$\times10^4$ events of $\Xi_{bb}$. There are only about $0.25\times 10^{2}$ events of $\Xi_{cc}$, $3.85\times 10^{2}$ events of $\Xi_{bc}$ and $0.17\times 10^{2}$ events of $\Xi_{bb}$ produced at the CEPC/ILC, where the production of doubly heavy baryons is accompanied by a cleaner background, which facilitates experimental measurements.
Considering a upgrade CEPC/ILC with the luminosity up to the level of HL-LHC, such as $3~ab^{-1}$, the produced envents of corresponding $\Xi_{QQ'}$ would increase about 3.75 times.
\end{itemize}

Theoretically, some phenomenological models have been proposed to predict the decay channels and discovery potential of the doubly heavy baryons~\cite{Ali:2018ifm,Ali:2018xfq,Qin:2020zlg}. An overview about the decay and observation of the doubly heavy baryons may be found in Refs.~\cite{Bediaga:2012py, Bediaga:2018lhg} by the LHCb collaboration. Similar to the observation of $\Xi_{cc}^{++}$, the produced $\Xi_{bc}$ and $\Xi_{bb}$ could be observed by cascade decays, such as $\Xi_{bc}^{+}\to\Xi_{cc}^{++}(\to p K^-\pi^+ \pi^+)\pi^{-}$ and $\Xi^{0}_{bb}\to\Xi_{bc}^+ (\to\Xi^{++}\pi^-) \pi^-$. As for the detection efficiency in experiment, the events can't be 100$\%$ detected. Despite the experimental limitations, optimistically, some signals through Higgs decays would still be likely to be detected in experiment with such a large base of $\Xi_{QQ'}$ production.

\subsection{Differential distributions}
\begin{figure}[htb]
  \centering
  \includegraphics[width=0.32\textwidth]{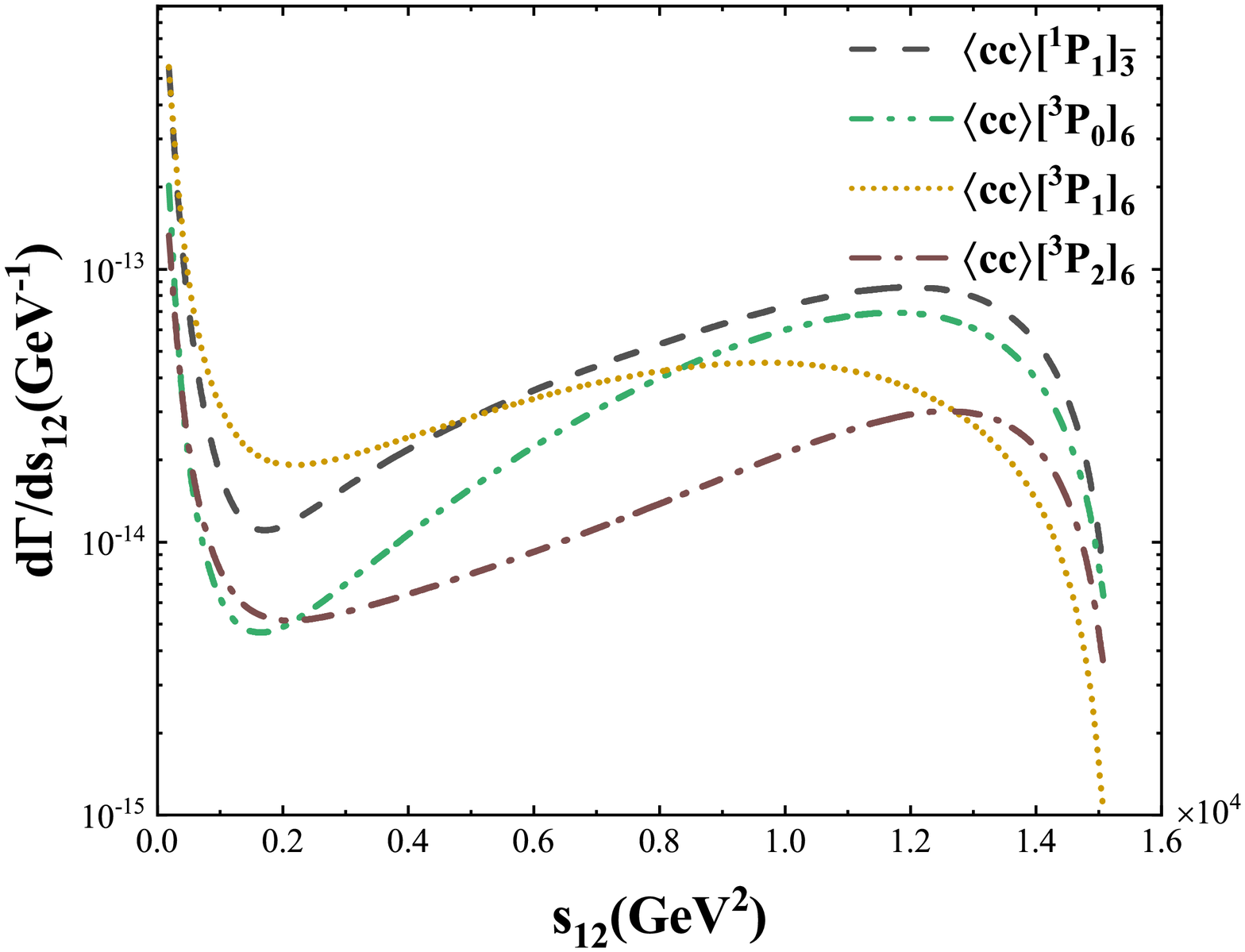}
  \includegraphics[width=0.32\textwidth]{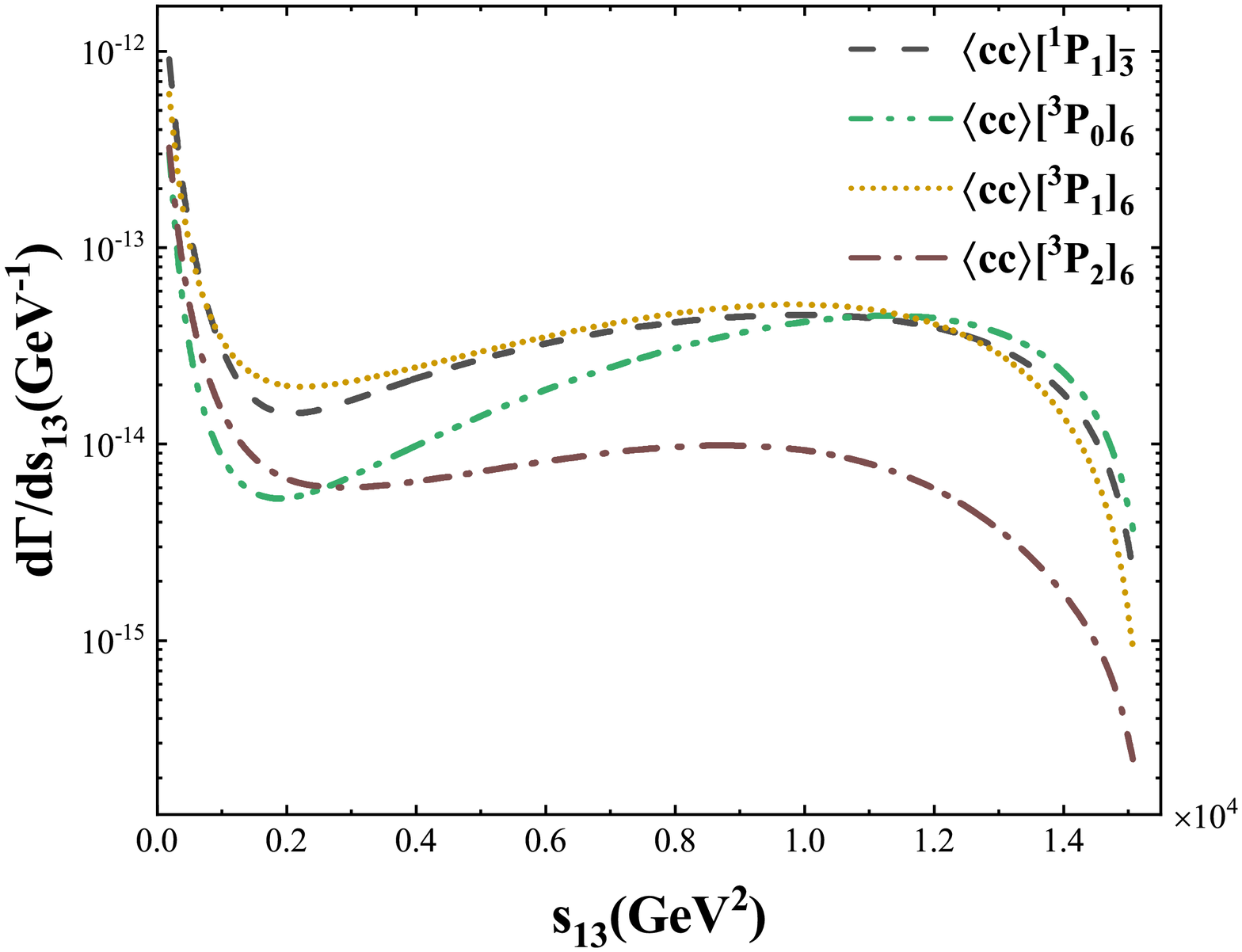}
  \includegraphics[width=0.32\textwidth]{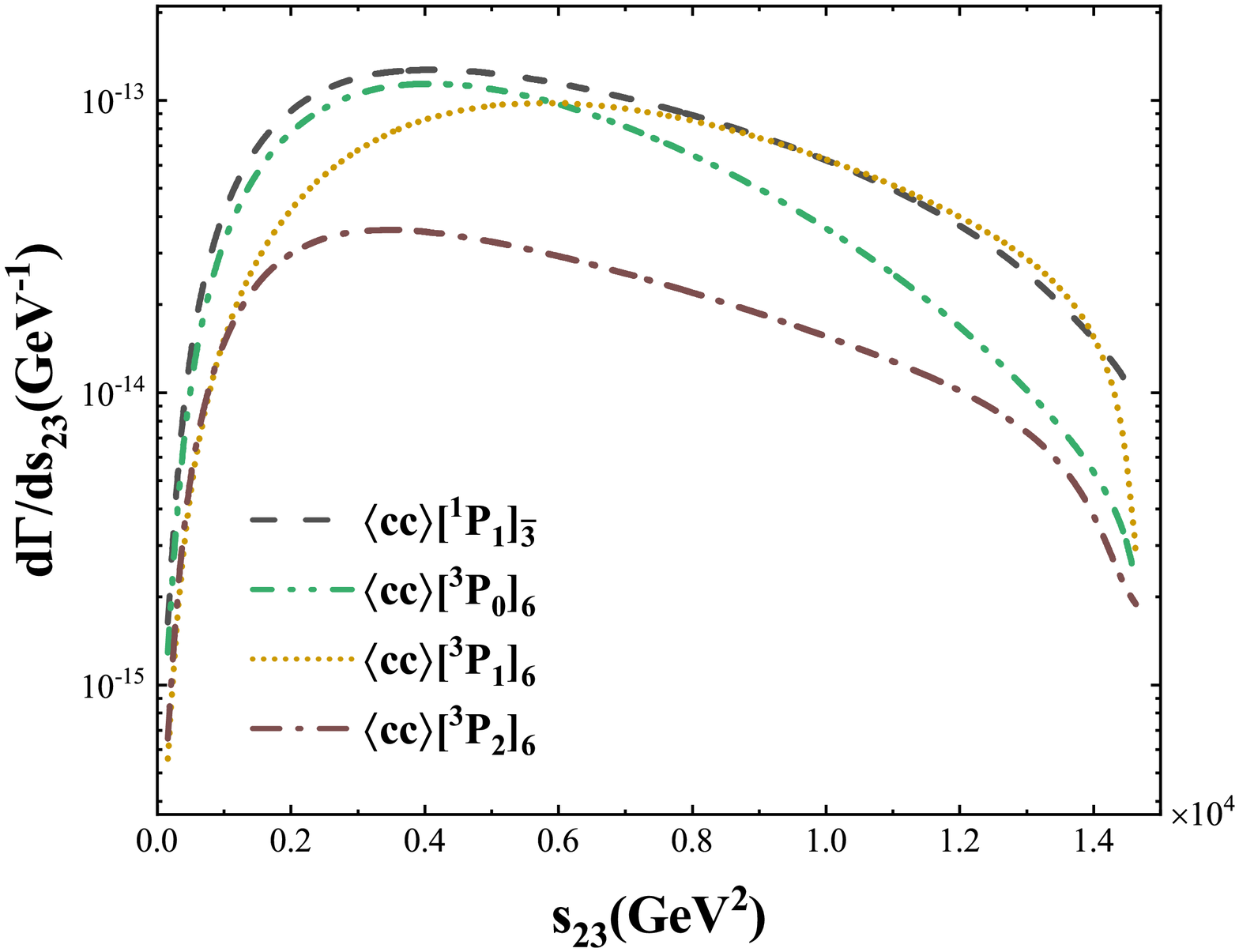}\\
  \includegraphics[width=0.32\textwidth]{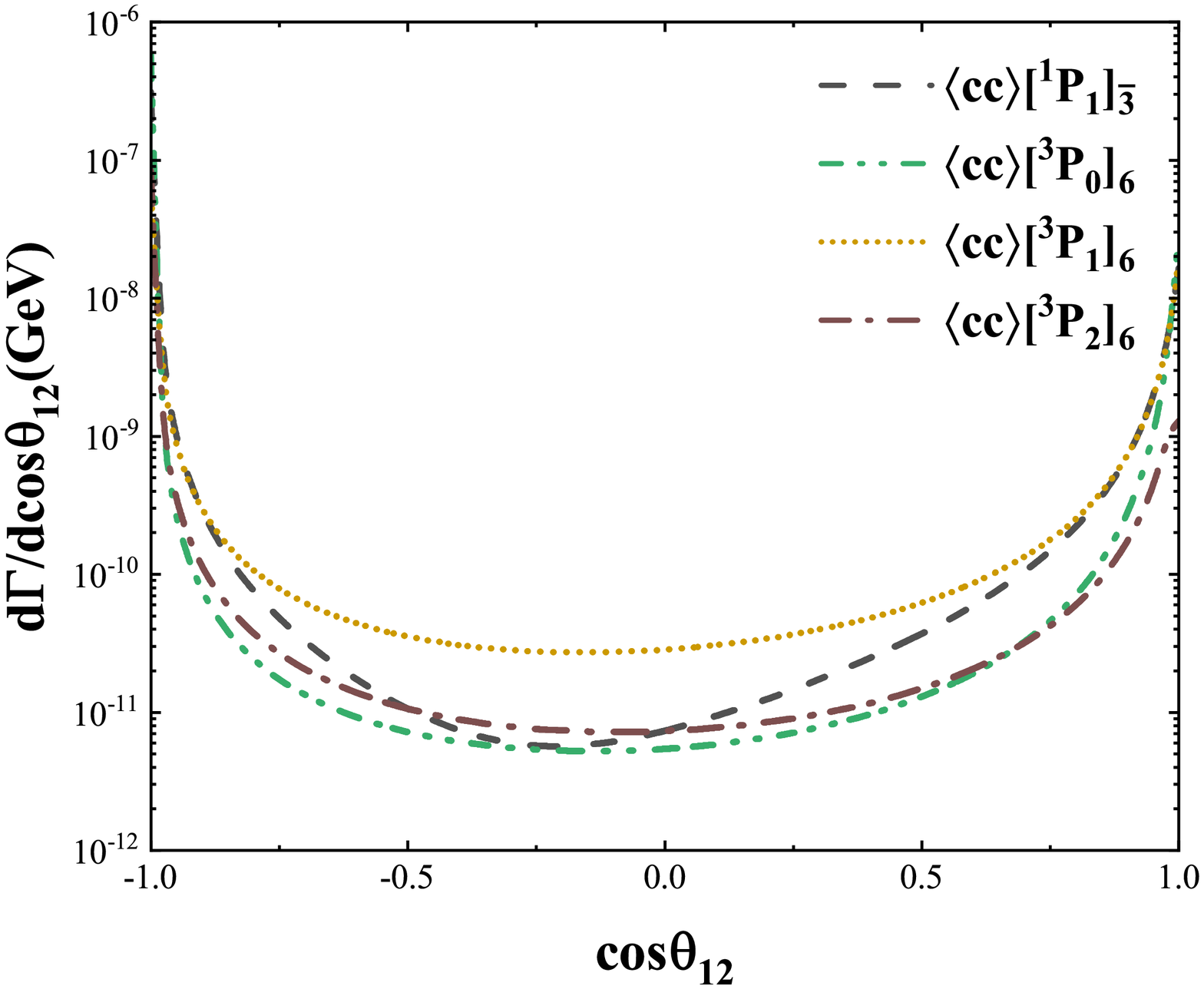}  
  \includegraphics[width=0.32\textwidth]{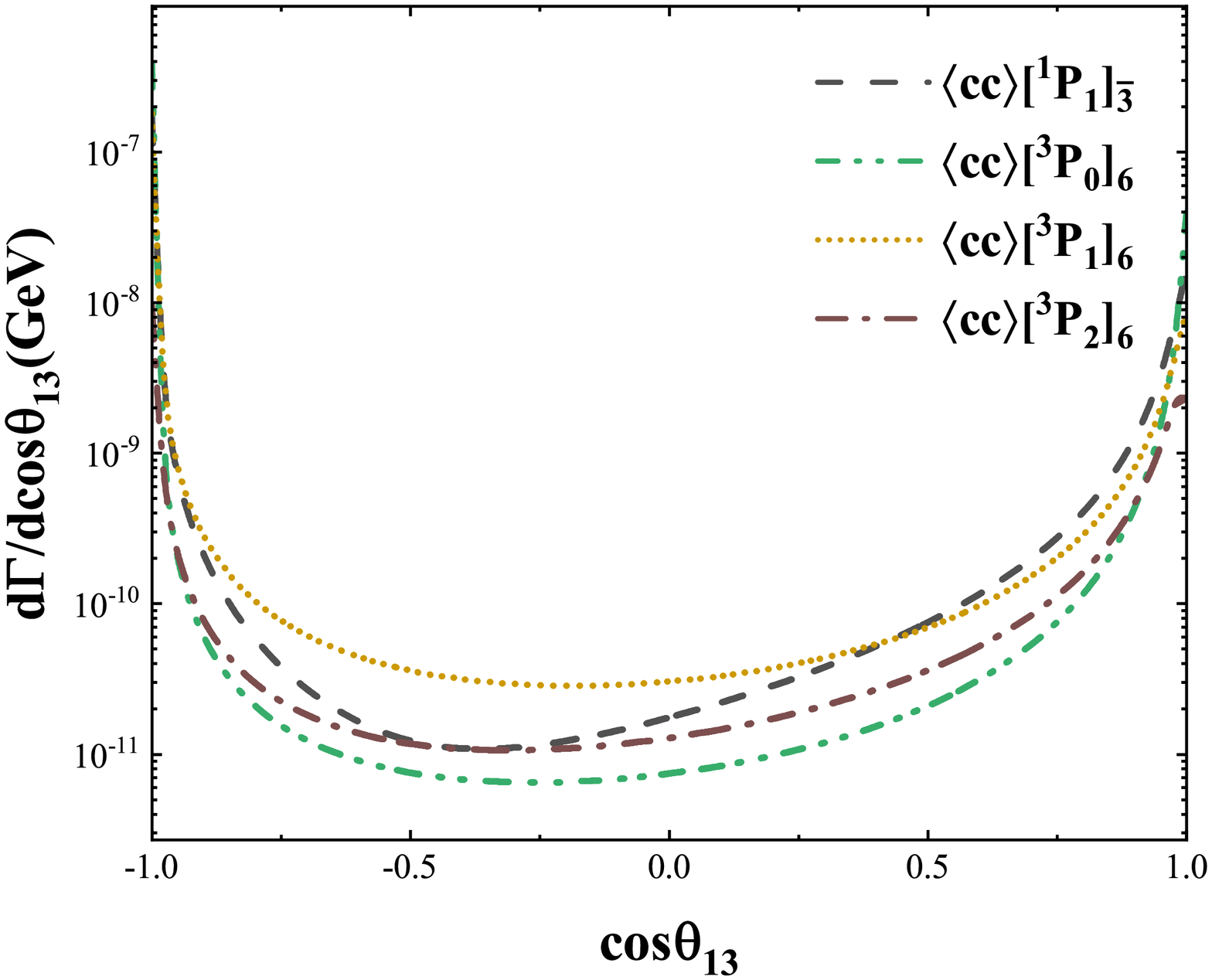}
  \includegraphics[width=0.32\textwidth]{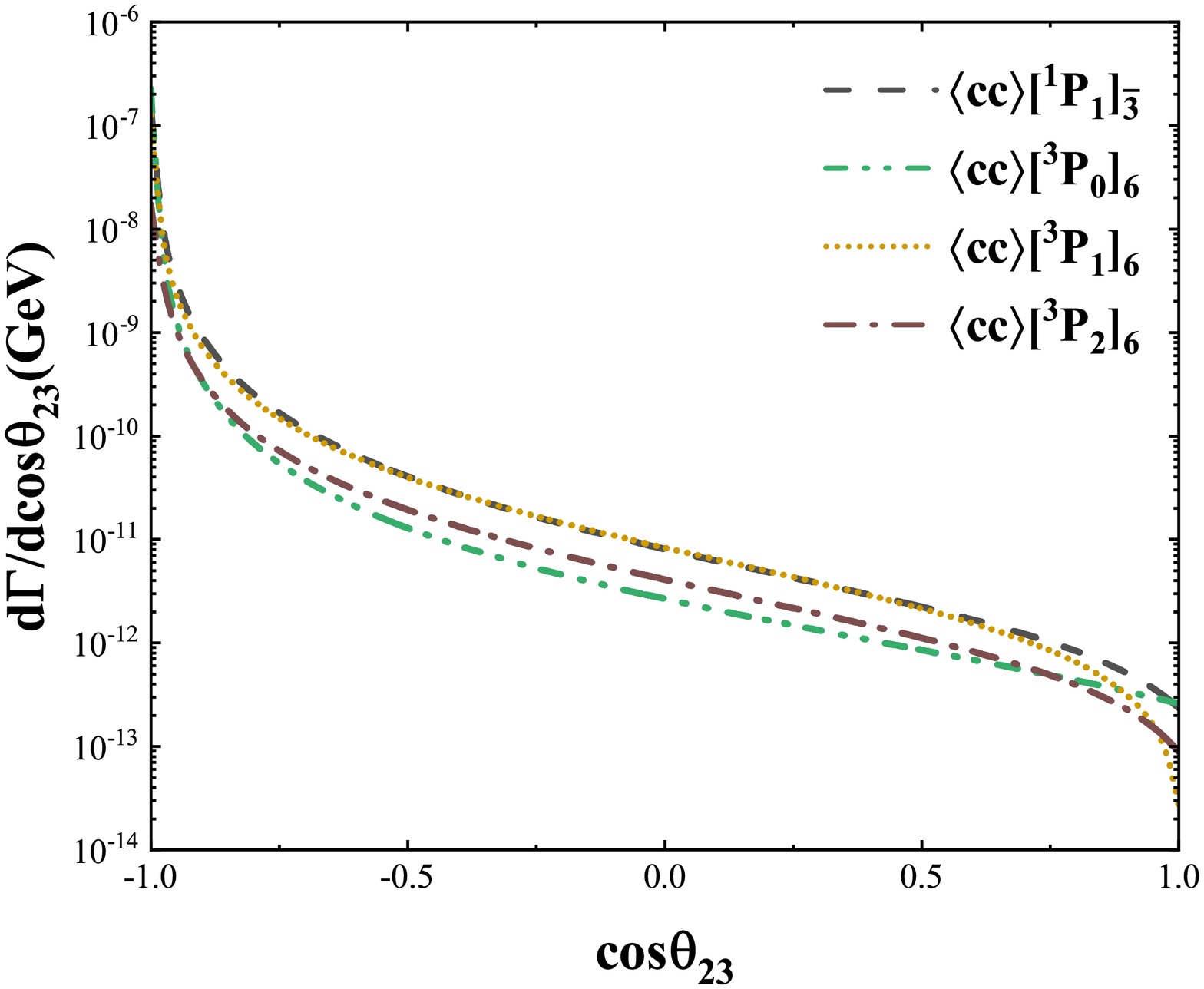}
  \caption{The invariant mass differential distributions $d\Gamma/d \rm s_{12}$, $d\Gamma/d \rm s_{13}$, $d\Gamma/d \rm s_{23}$ and the angular differential distributions $d\Gamma/ d\rm cos\theta_{12}$, $d\Gamma/ d\rm cos\theta_{13}$, $d\Gamma/ d\rm cos\theta_{23}$ for the process $H(p_0) \rightarrow \Xi_{cc}(p_1)+ \bar {c}(p_2) + \bar {c}(p_3)$. Four colorful lines represent the considered intermediate diquark states, i.e., $\langle cc\rangle[^{1}P_{1}]_{\mathbf{\bar 3}}$ and $\langle cc\rangle[^{3}P_{J}]_{\mathbf{6}}$ with $J$=0, 1, 2.}
  \label{cc} 
\end{figure}

\begin{figure}[htb]
  \centering
  \includegraphics[width=0.32\textwidth]{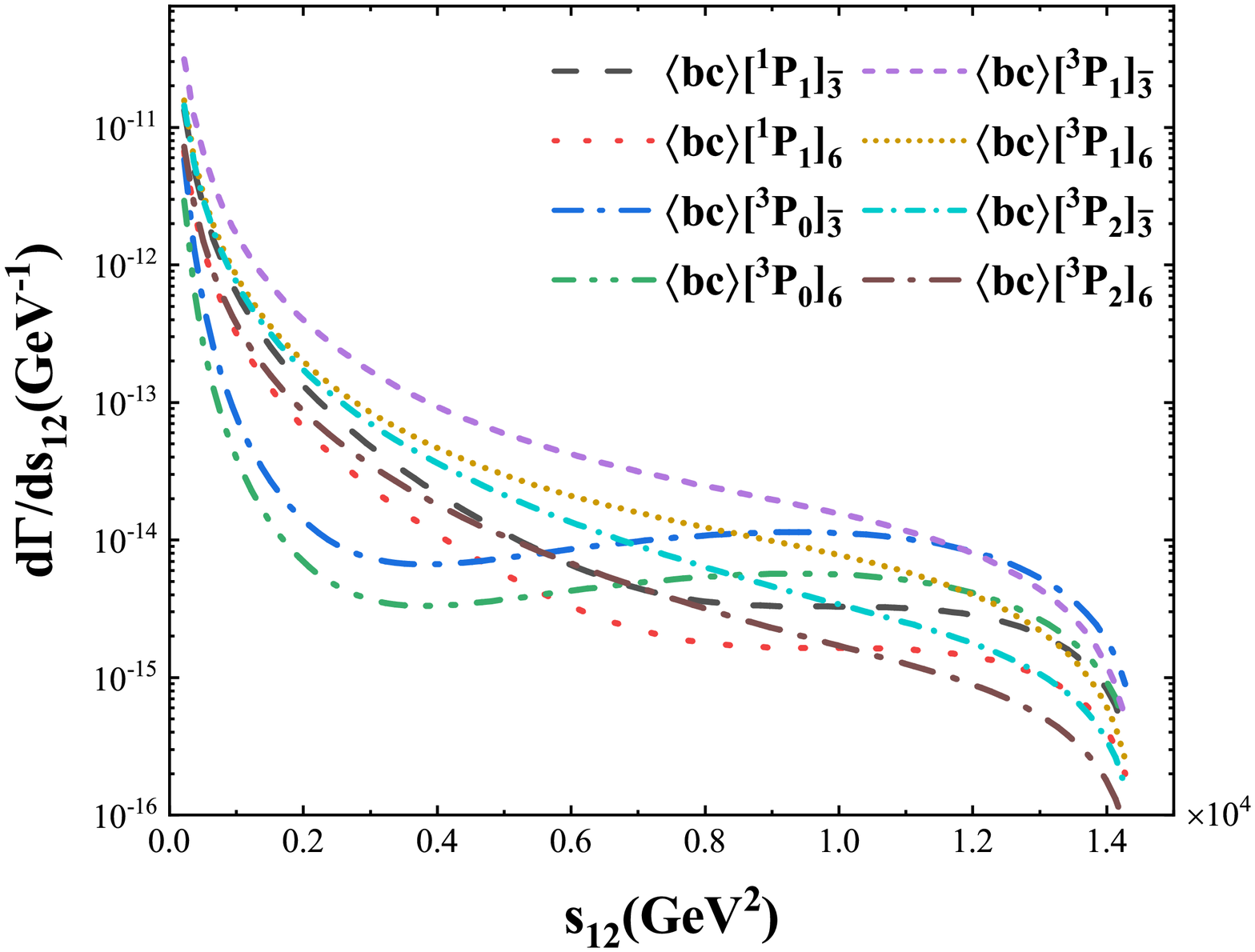}
  \includegraphics[width=0.32\textwidth]{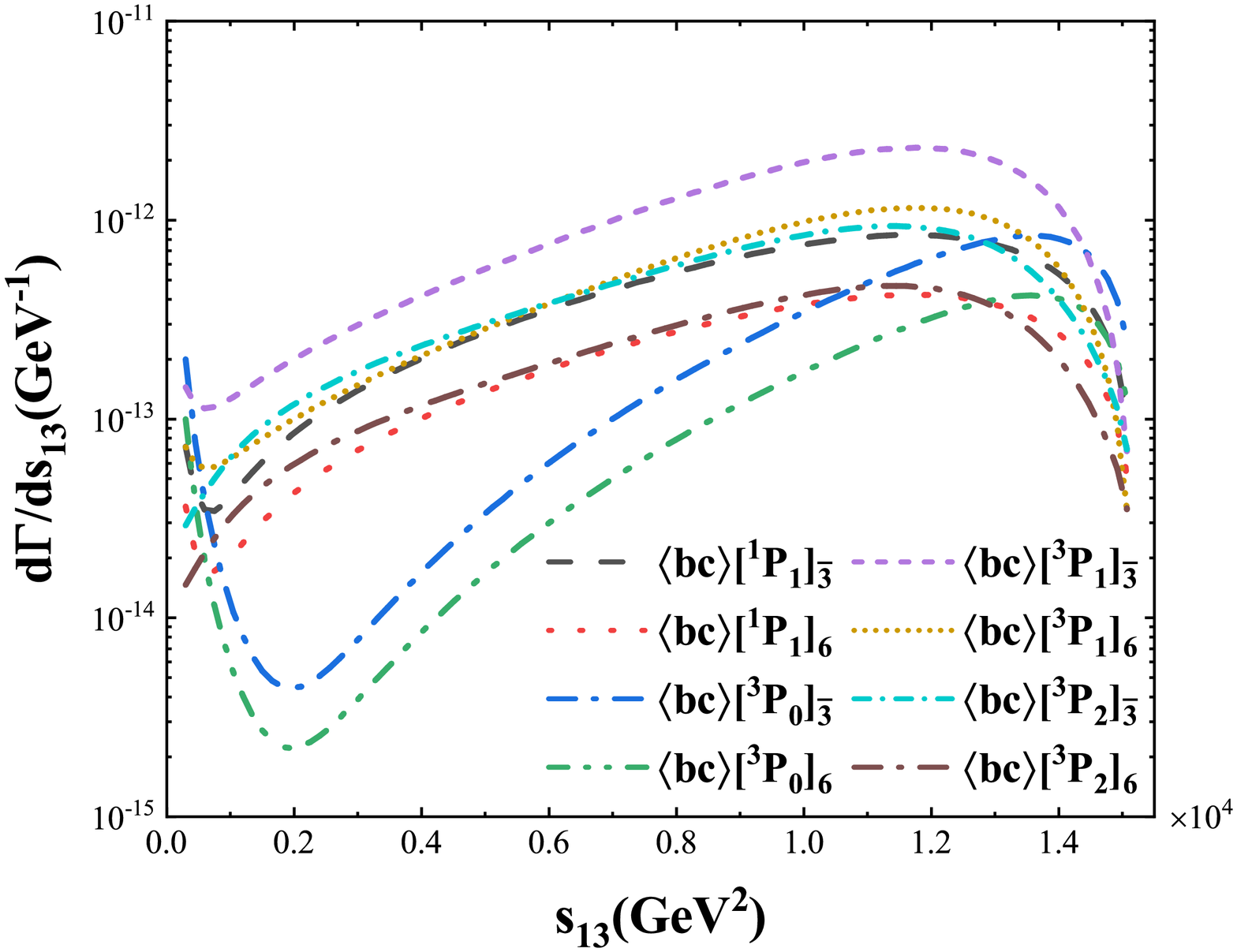}
  \includegraphics[width=0.32\textwidth]{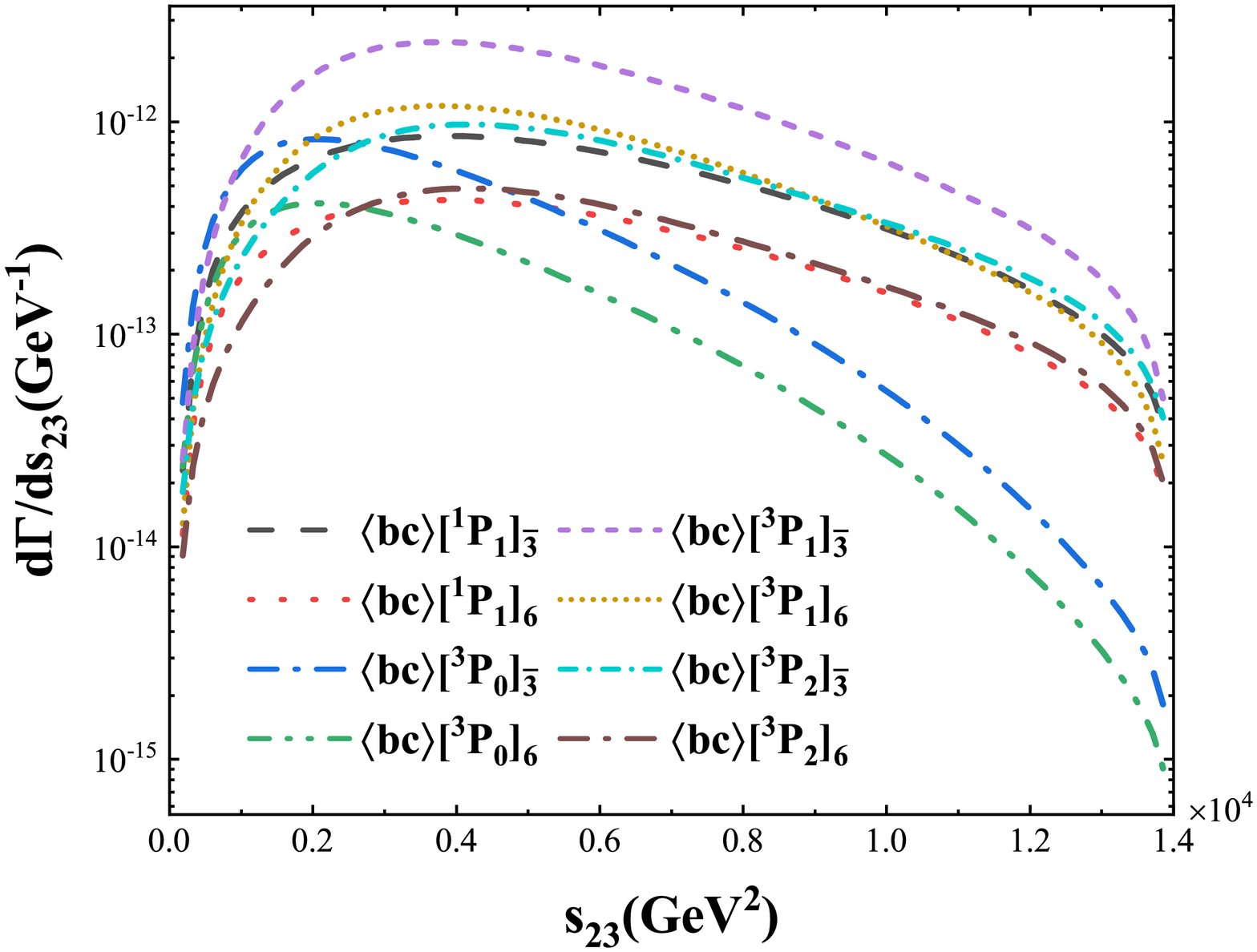}\\
  \includegraphics[width=0.32\textwidth]{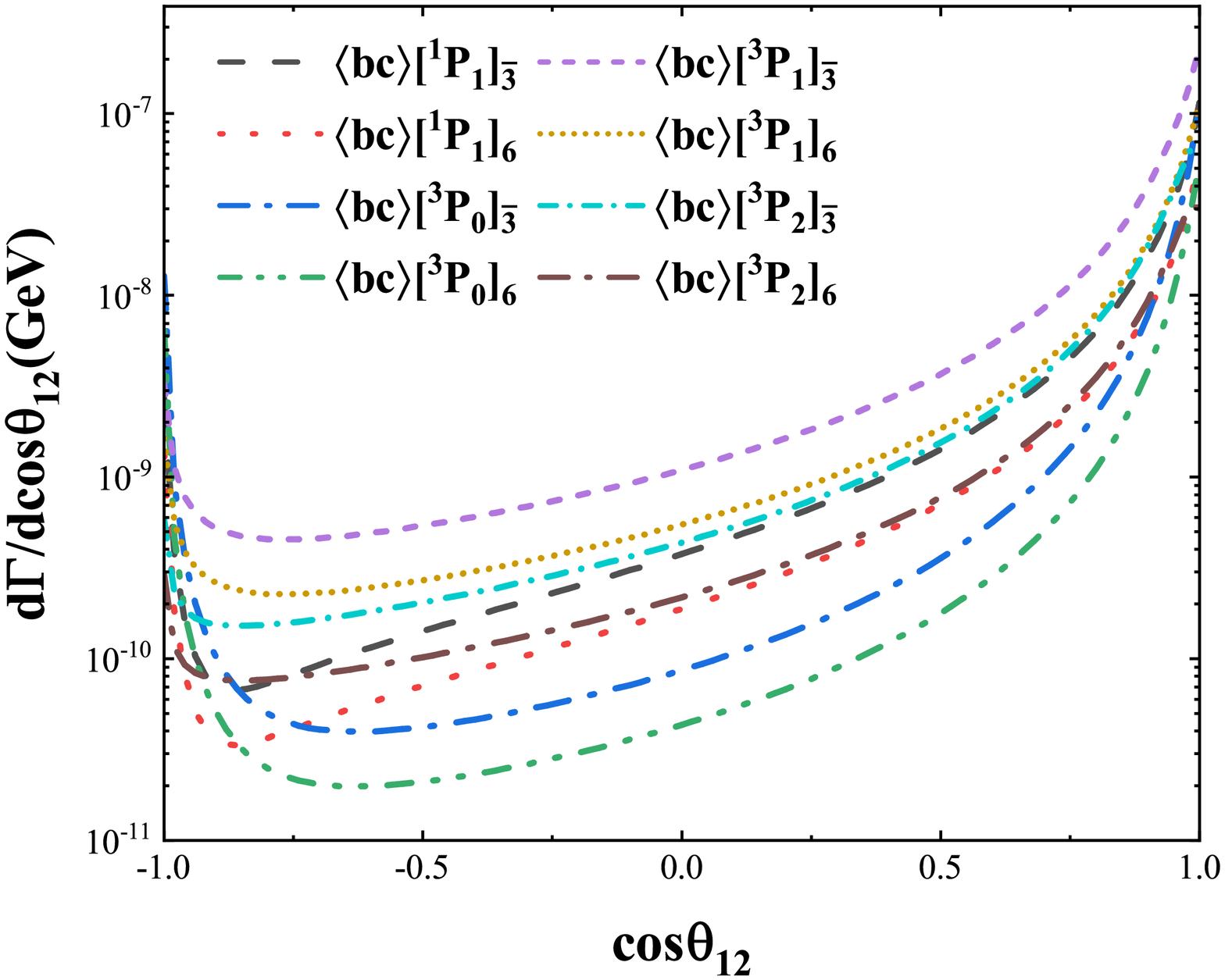}  
  \includegraphics[width=0.32\textwidth]{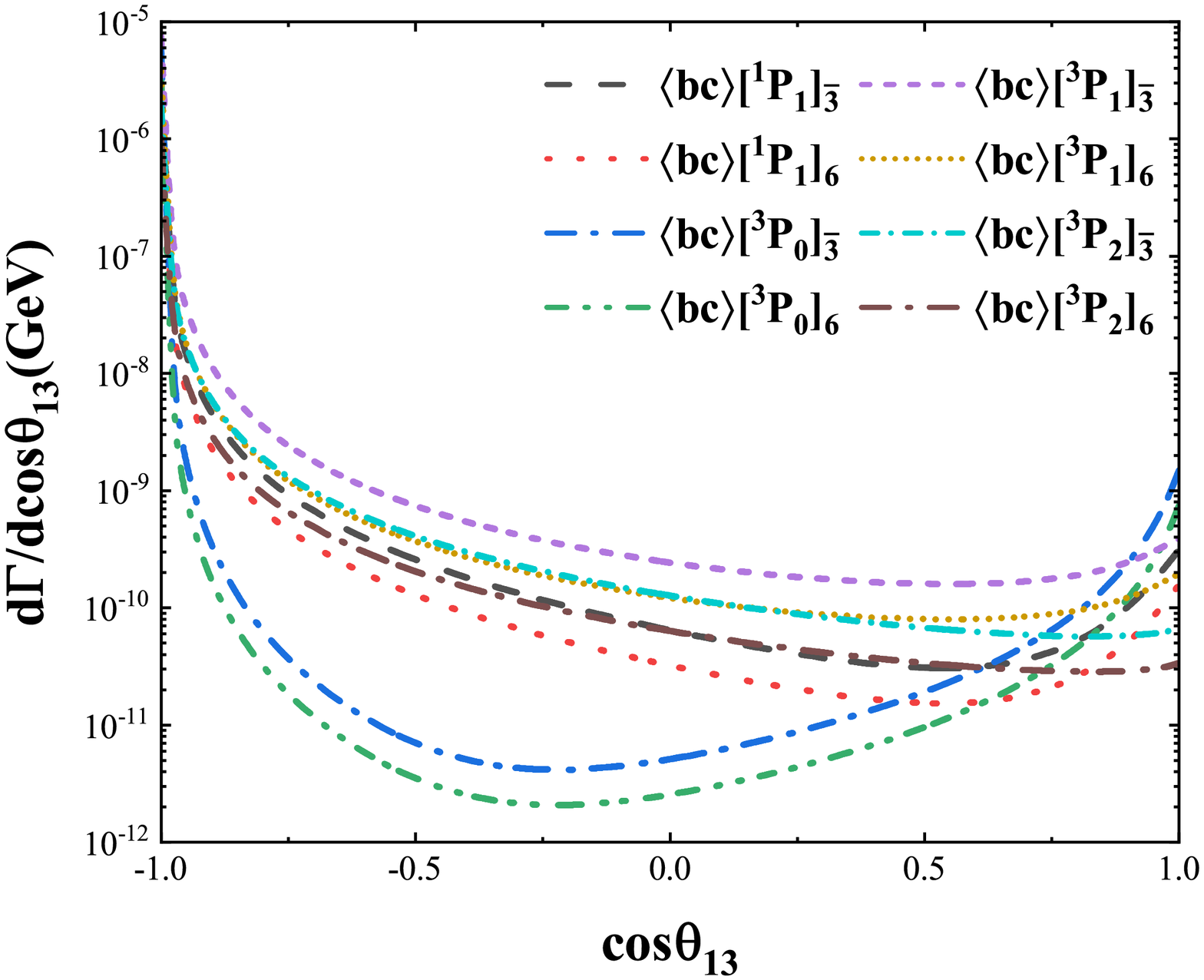}
  \includegraphics[width=0.32\textwidth]{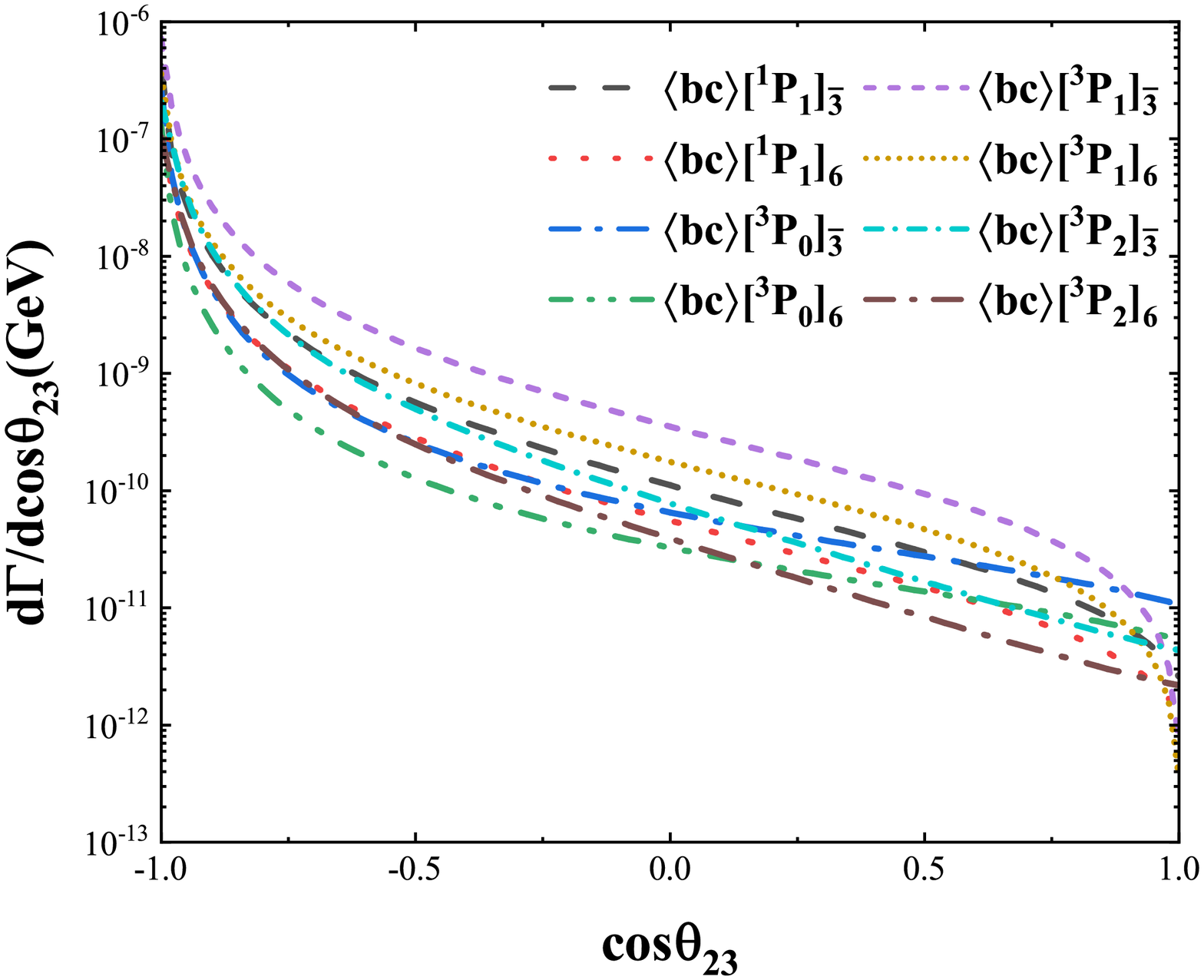}
  \caption{The invariant mass differential distributions  $d\Gamma/d \rm s_{12}$, $d\Gamma/d \rm s_{13}$, $d\Gamma/d \rm s_{23}$ and the angular differential distributions $d\Gamma/ d\rm cos\theta_{12}$, $d\Gamma/ d\rm cos\theta_{13}$, $d\Gamma/ d\rm cos\theta_{23}$ for the process $H(p_0) \rightarrow \Xi_{bc}(p_1)+ \bar {c}(p_2) + \bar {b}(p_3)$. Eight colorful lines represent the considered intermediate diquark states, i.e., $\langle bc\rangle[^{1}P_{1}]_{\mathbf{\bar 3}/ \mathbf{6}}$ and $\langle bc\rangle[^{3}P_{J}]_{\mathbf{\bar 3}/ \mathbf{6}}$ with $J$=0, 1, 2.}
  \label{bc} 
\end{figure}

\begin{figure}[htb]
  \centering
  \includegraphics[width=0.32\textwidth]{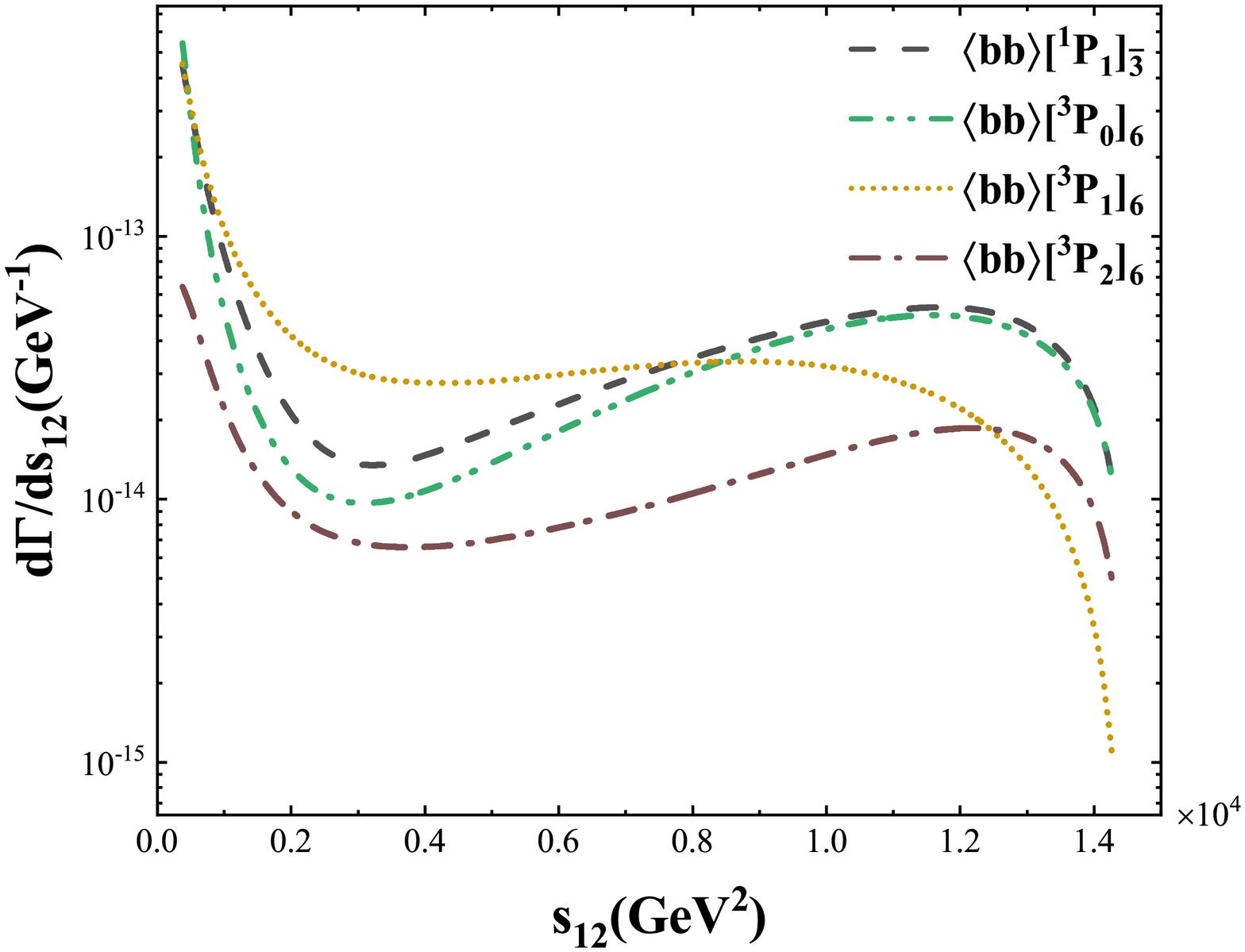}
  \includegraphics[width=0.32\textwidth]{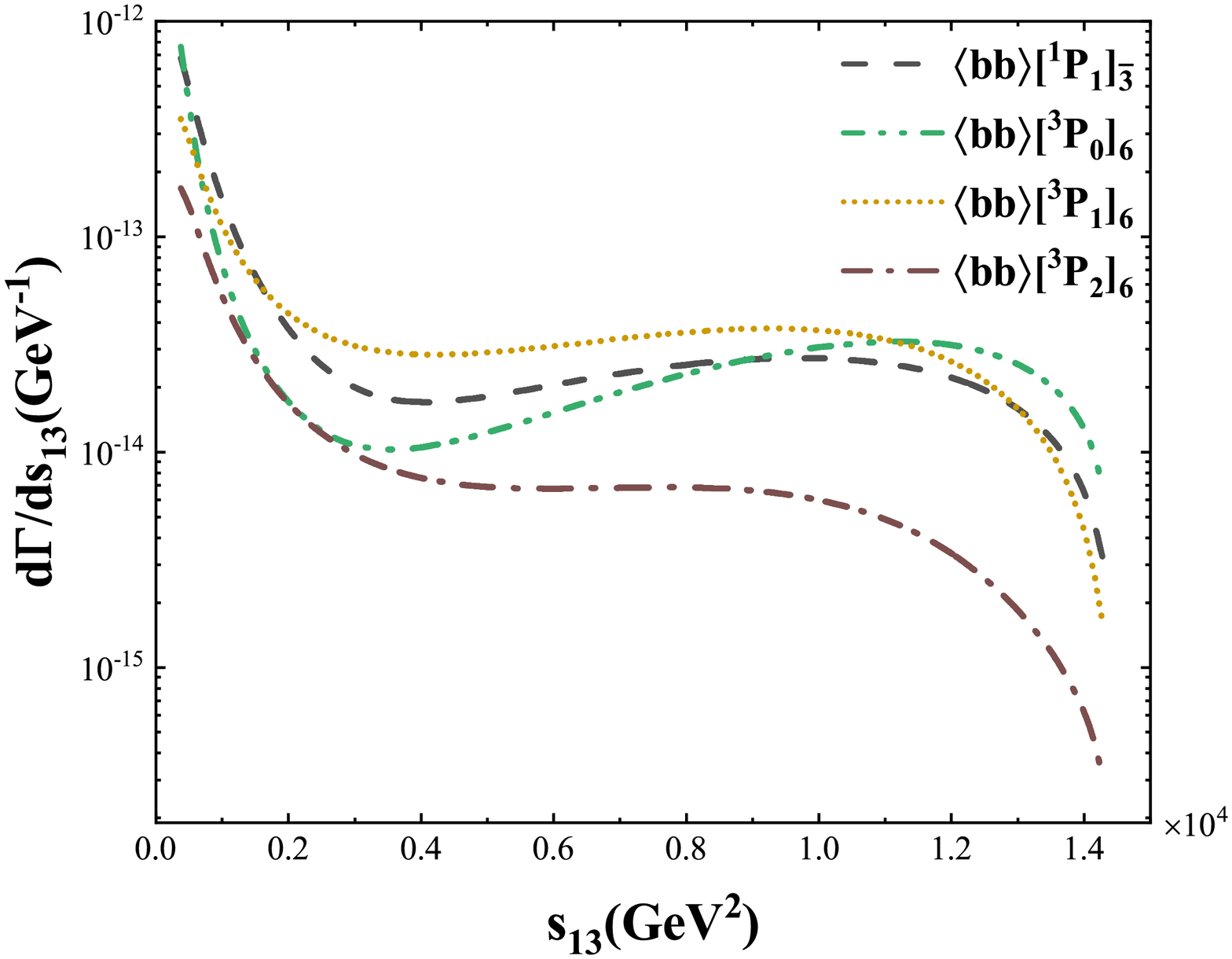}
  \includegraphics[width=0.32\textwidth]{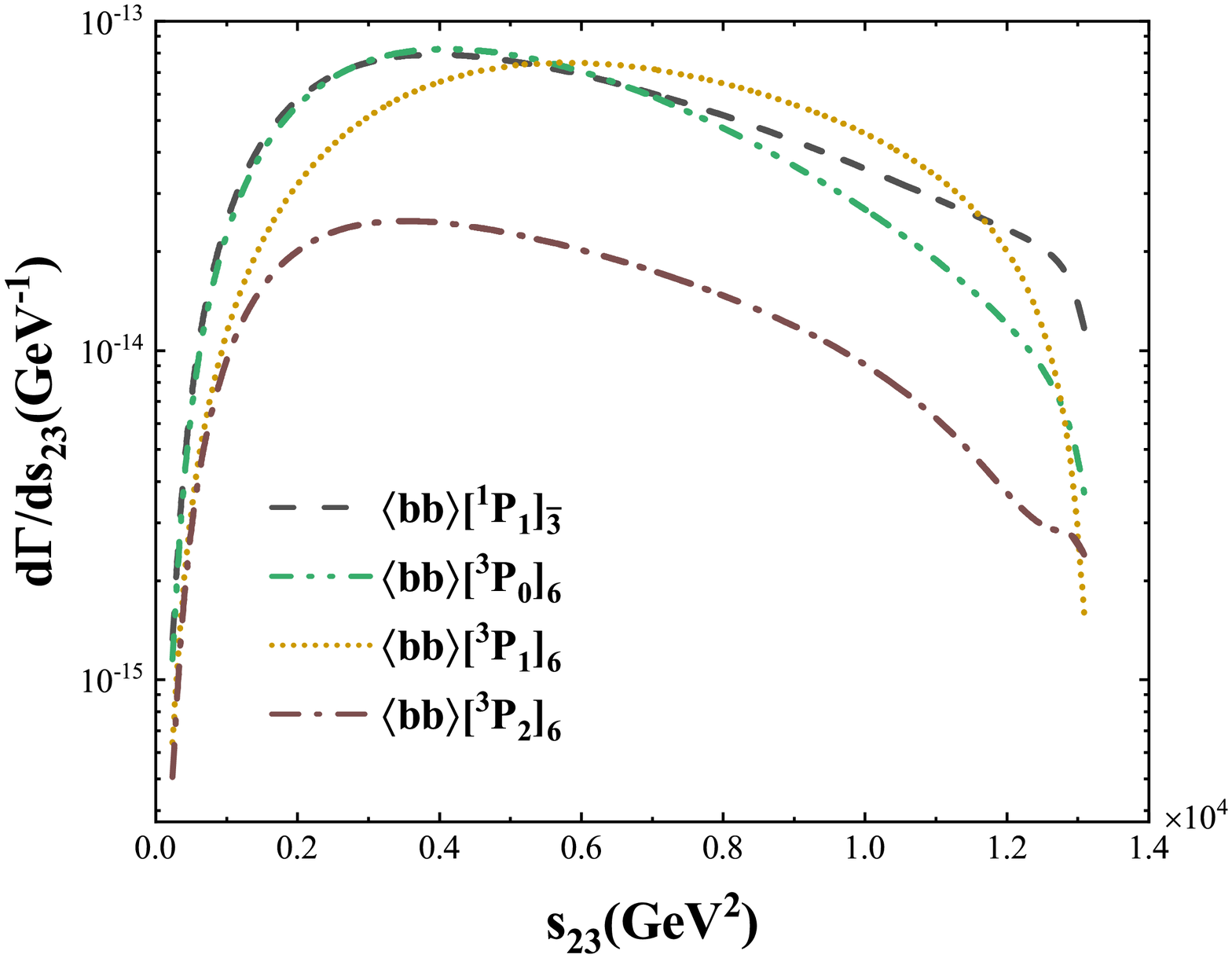}\\
  \includegraphics[width=0.32\textwidth]{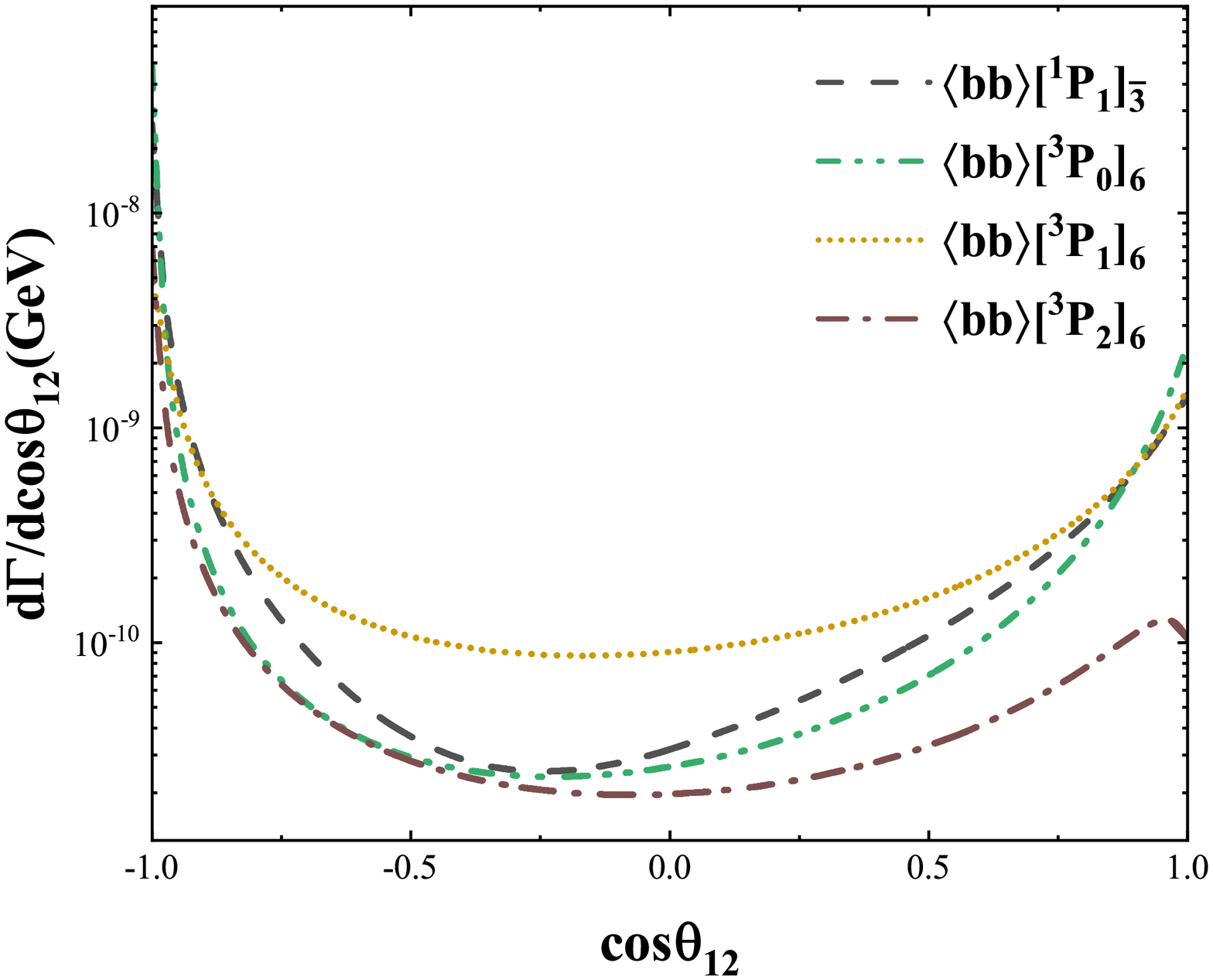}  
  \includegraphics[width=0.32\textwidth]{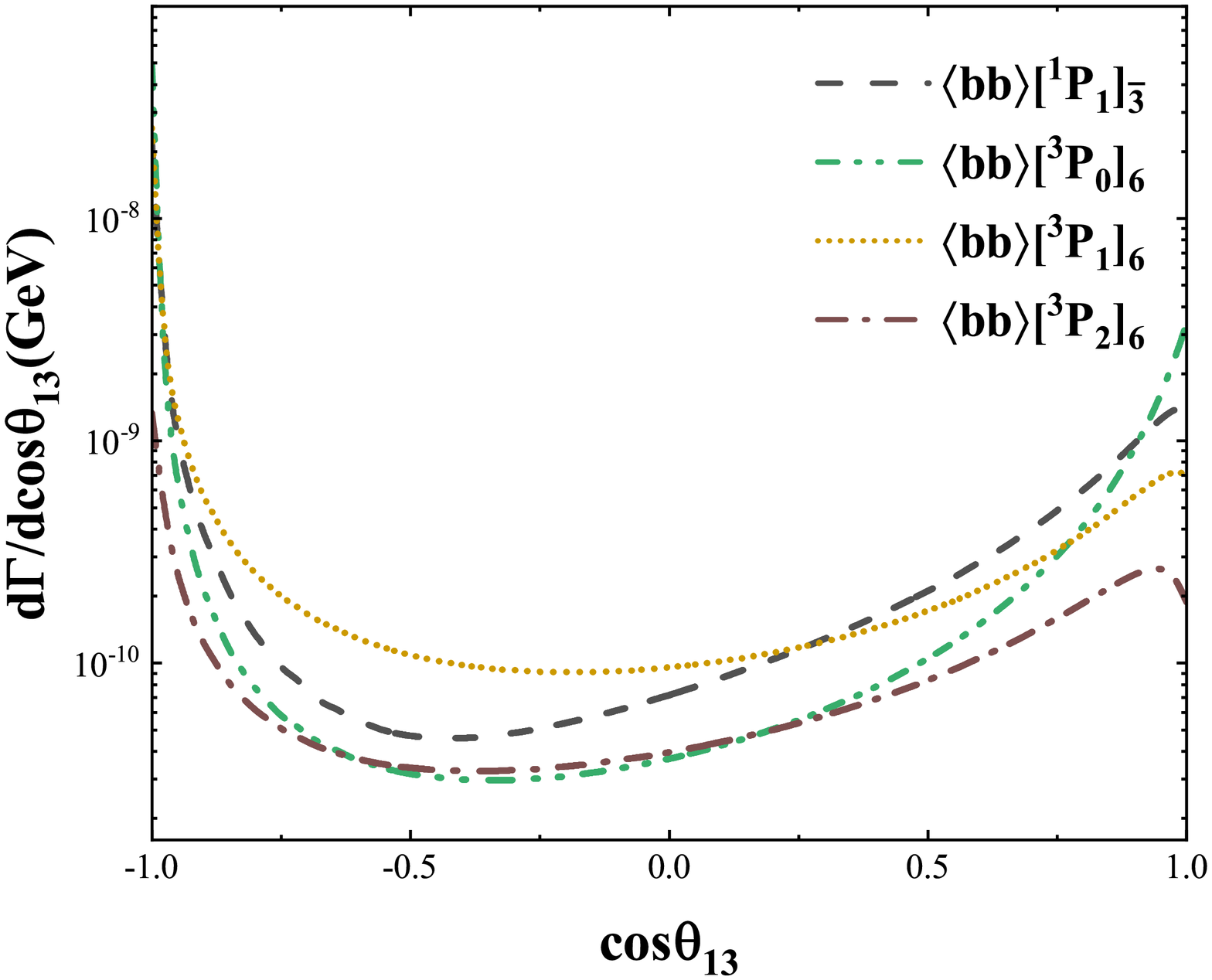}
  \includegraphics[width=0.32\textwidth]{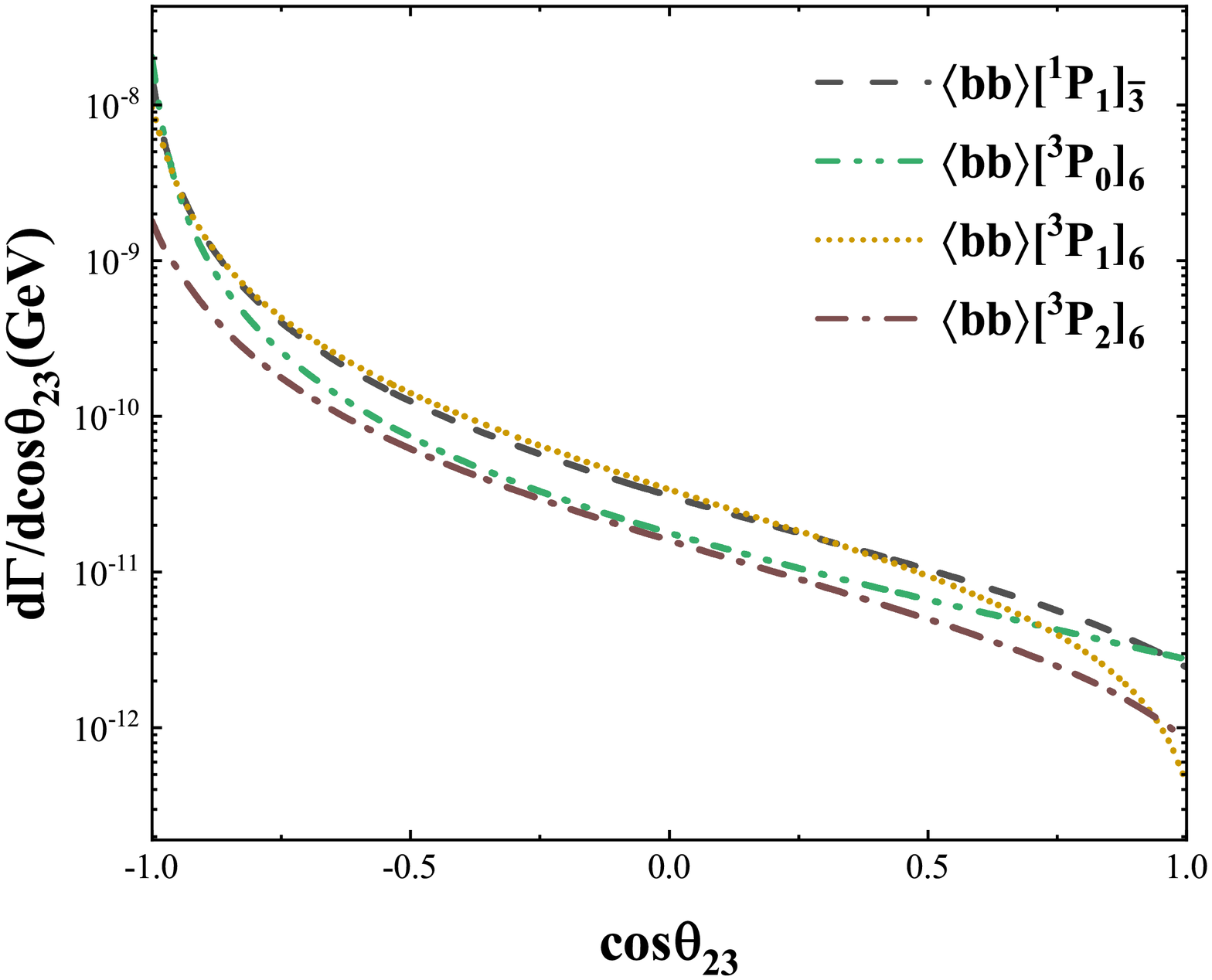}
  \caption{The invariant mass differential distributions $d\Gamma/d \rm s_{12}$, $d\Gamma/d \rm s_{13}$, $d\Gamma/d \rm s_{23}$ and the angular differential distributions $d\Gamma/ d\rm cos\theta_{12}$, $d\Gamma/ d\rm cos\theta_{13}$, $d\Gamma/ d\rm cos\theta_{23}$ for the process $H(p_0) \rightarrow \Xi_{bb}(p_1)+ \bar {b}(p_2) + \bar {b}(p_3)$. Four colorful lines represent the considered intermediate diquark states, i.e.,
$\langle bb\rangle[^{1}P_{1}]_{\mathbf{\bar 3}}$ and $\langle bb\rangle[^{3}P_{J}]_{\mathbf{6}}$ with $J$=0, 1, 2.}
  \label{bb} 
\end{figure}

In order to analyze the kinematics of this process in which the doubly heavy baryons $\Xi_{QQ'}$ are indirected production via Higgs decays and give some guidance in the experimental search for the doubly heavy baryons, we present the invariant mass differential distributions $d\Gamma/d \rm s_{ij}$ and the angular differential distributions $d\Gamma/ d\rm cos\theta_{ij}$ in Figs.~\ref{cc},~\ref{bc} and \ref{bb} for the production of $\Xi_{cc}$, $\Xi_{bc}$, and $\Xi_{bb}$ accordingly, where the invariant mass $\rm s_{ij}=\it (p_{i}+p_{j})^{\rm 2}$ and $\theta_{ij}$ is the angle between the momenta $\overrightarrow{p_{i}}$ and $\overrightarrow{p_{j}}$ in the Higgs boson rest frame.
All the contributions of considered intermediate diquark states have been displayed into the corresponding Figs.~(\ref{cc}-\ref{bb}).


From Figs.~(\ref{cc}-\ref{bb}), we can see that the behaviors of the $P$-wave differential distributions, including the invariant mass and the angular differential distributions, are similar to that of the $S$-wave $\Xi_{QQ'}$ production through the same process~\cite{Niu:2019xuq}. 
And for the angular differential decay widths $d\Gamma/ d\rm cos\theta_{23}$ in Figs.(\ref{cc}-\ref{bb}), the largest contribution can be obtained when $\rm cos\theta_{23}=-1~(\theta_{23}=\pi$), which means that the two antiquarks in the final state always move back to back. The invariant mass differential distribution $d\Gamma/ d\rm s_{12}$ could reach a maximum when the invariant mass $\rm s_{12}$ is at its minimum value.

For the production of $P$-wave $\Xi_{cc}$ and $\Xi_{bb}$ in Figs.~\ref{cc} and \ref{bb}, all the kinematic features are the same, except that the differential decay width of $\Xi_{cc}$ is slightly larger than that of $\Xi_{bb}$. The differential distributions of the invariant mass $\rm s_{13}$ is similar to that of $\rm s_{12}$. And the angular differential distributions of $\cos\theta_{12}$ and $\cos\theta_{13}$ is not very distinct for the doubly heavy baryons produced with two identical anti-particles.

While for the excited $\Xi_{bc}$ production in Fig.~\ref{bc}, we can see that the behaviors of the diquark states $\langle bc\rangle[^{3}P_{0}]_{\mathbf{\bar{3}/6}}$ change more dramatically than that of the other diquark states in the differential distributions. And the differential decay widths in the three angular differential distributions of Fig.~\ref{bc} could achieve its biggest when the $P$-wave $\Xi_{bc}$ move in the same direction as $\bar{c}$ and the opposite direction as $\bar{b}$ for $\cos\theta_{12}=1$, $\cos\theta_{13}=-1$ and $\cos\theta_{23}=-1$~($\theta_{12}=0,~\theta_{13}=\pi,~\theta_{23}=\pi$).

\subsection{Theoretical uncertainty}

For the production of excited $\Xi_{QQ'}$ through the process $H \rightarrow \Xi_{QQ'}+ \bar {Q'} + \bar {Q}$, the main theoretical uncertainty arises from the mass of heavy quarks, as the mass of heavy quarks affects not only the mass of doubly heavy baryons, but also the phase space of this process. In the following, we would vary $m_c=1.8\pm0.3~\rm{GeV}$ to analyze the uncertainty for the excited $\Xi_{cc}$ production and $m_b=5.1\pm0.4~\rm{GeV}$ for the excited $\Xi_{bb}$ production. As for the production of $\Xi_{bc}$, the results are related to both $m_c$ and $m_b$. We would study the influence of $m_c$ and $m_b$ on the results separately, keeping the other parameters at the central values.

The theoretical uncertainty caused by the heavy quark mass $m_c$ and $m_b$ for the excited $\Xi_{QQ'}$ are presented in Table~\ref{mc} and \ref{mb}, respectively.
From Table~\ref{mc}, we can find that for the suppression of phase space, the decay widths of $P$-wave $\Xi_{cc}$ and $\Xi_{bc}$ both decrease with the increment of $m_c$. And $m_c$ has a greater impact on the decay widths of $\Xi_{bc}$ than that of $\Xi_{cc}$ for the Yukawa coupling.
Table~\ref{mb} shows that for the production of excited $\Xi_{bc}$, the decay widths increase with the increment of $m_b$. That is because with the increase of $m_b$, the Yukawa coupling increases while the phase space is suppressed. Moreover, the Yukawa coupling has a greater influence on the decay widths and dominates the decay widths increase. Like the trend of $\Xi_{cc}$ toward $m_c$, the decay widths for the production of $\Xi_{bb}$ also decreases with the increment of $m_b$ according to the suppression of phase space.

\begin{table}[htb]
\begin{center}
\caption{Theoretical uncertainties for the production of $P$-wave $\Xi_{cc}$ and $\Xi_{bc}$ by varying $m_c=1.8 \pm 0.3~\rm{GeV}$. The units of the decay widths are $10^{-9}$~GeV.} \vspace{0.5cm}
\begin{tabular}{|c|c|c|c|c|c|} 
			\hline
			$m_c$(GeV)                                           & 1.50~  & 1.65~  & 1.80~  & 1.95~  & 2.10~   \\ 
			\hline\hline
			$\Gamma_{\Xi_{cc}}([^1P_1]_{\overline{\mathbf{3}}})$ & 1.89~  & 1.41~  & 1.08~  & 0.84~  & 0.67~   \\ 
			\hline
			$\Gamma_{\Xi_{cc}}([^3P_0]_{\mathbf{6}})$            & 1.43~  & 1.07~  & 0.83~  & 0.65~  & 0.52~   \\ 
			\hline
			$\Gamma_{\Xi_{cc}}([^3P_1]_{\mathbf{6}})$            & 1.47~  & 1.11~  & 0.86~  & 0.67~  & 0.54~   \\ 
			\hline
			$\Gamma_{\Xi_{cc}}([^3P_2]_{\mathbf{6}})$            & 0.51~  & 0.38~  & 0.29~  & 0.23~  & 0.18~   \\
			\hline\hline
			$\Gamma_{\Xi_{bc}}([^1P_1]_{\overline{\mathbf{3}}})$ & 15.95~ & 10.03~ & 6.60~  & 4.51~  & 3.19~   \\ 
			\hline
			$\Gamma_{\Xi_{bc}}([^1P_1]_{\mathbf{6}})$            & 7.97~  & 5.02~  & 3.30~  & 2.26~  & 1.59~   \\ 
			\hline
			$\Gamma_{\Xi_{bc}}([^3P_0]_{\overline{\mathbf{3}}})$ & 8.01~  & 5.59~  & 4.07~  & 3.06~  & 2.37~   \\ 
			\hline
			$\Gamma_{\Xi_{bc}}([^3P_0]_{\mathbf{6}})$            & 4.01~  & 2.80~  & 2.03~  & 1.53~  & 1.19~   \\ 
			\hline
			$\Gamma_{\Xi_{bc}}([^3P_1]_{\overline{\mathbf{3}}})$ & 40.21~ & 24.82~ & 16.07~ & 10.84~ & 7.57~   \\ 
			\hline
			$\Gamma_{\Xi_{bc}}([^3P_1]_{\mathbf{6}})$            & 20.11~ & 12.41~ & 8.04~  & 5.42~  & 3.78~   \\ 
			\hline
			$\Gamma_{\Xi_{bc}}([^3P_2]_{\overline{\mathbf{3}}})$ & 18.63~ & 11.12~ & 6.95~  & 4.51~  & 3.02~   \\ 
			\hline
			$\Gamma_{\Xi_{bc}}([^3P_2]_{\mathbf{6}})$            & 9.31~  & 5.56~  & 3.47~  & 2.25~  & 1.51~   \\ 
			\hline
		\end{tabular}
\label{mc}
\end{center}
\end{table}

\begin{table}[htb]
\begin{center}
\caption{Theoretical uncertainties for the production of $P$-wave $\Xi_{bc}$ and $\Xi_{bb}$ by varying $m_b=5.1 \pm 0.4~\rm{GeV}$. The units of the decay widths are $10^{-9}$~GeV.} \vspace{0.5cm}
\begin{tabular}{|c|c|c|c|c|c|} 
\hline
$m_b$(GeV)                                           & 4.7    & 4.9    & 5.1    & 5.3    & 5.5     \\ 
\hline\hline
$\Gamma_{\Xi_{bc}}([^1P_1]_{\overline{\mathbf{3}}})$ & 5.79~  & 6.19~  & 6.60~  & 7.04~  & 7.49~  \\ 
\hline
$\Gamma_{\Xi_{bc}}([^1P_1]_{\mathbf{6}})$            & 2.90~  & 3.09~  & 3.30~  & 3.52~  & 3.74~  \\ 
\hline
$\Gamma_{\Xi_{bc}}([^3P_0]_{\overline{\mathbf{3}}})$ & 3.90~  & 3.98~  & 4.07~  & 4.16~  & 4.26~  \\ 
\hline
$\Gamma_{\Xi_{bc}}([^3P_0]_{\mathbf{6}})$            & 1.95~  & 1.99~  & 2.03~  & 2.08~  & 2.13~   \\ 
\hline
$\Gamma_{\Xi_{bc}}([^3P_1]_{\overline{\mathbf{3}}})$ & 13.84~ & 14.93~ & 16.07~ & 17.27~ & 18.53~  \\ 
\hline
$\Gamma_{\Xi_{bc}}([^3P_1]_{\mathbf{6}})$            & 6.92~  & 7.46~  & 8.04~  & 8.64~  & 9.26~  \\ 
\hline
$\Gamma_{\Xi_{bc}}([^3P_2]_{\overline{\mathbf{3}}})$ & 5.76~  & 6.34~  & 6.95~  & 7.59~  & 8.26~  \\ 
\hline
$\Gamma_{\Xi_{bc}}([^3P_2]_{\mathbf{6}})$            & 2.88~  & 3.17~  & 3.47~  & 3.79~  & 4.13   \\ 
\hline\hline
$\Gamma_{\Xi_{bb}}([^1P_1]_{\overline{\mathbf{3}}})$ & 0.62~  & 0.54~  & 0.47~  & 0.42~  & 0.37~   \\ 
\hline
$\Gamma_{\Xi_{bb}}([^3P_0]_{\mathbf{6}})$            & 0.56~  & 0.50~  & 0.44~  & 0.39~  & 0.35~   \\ 
\hline
$\Gamma_{\Xi_{bb}}([^3P_1]_{\mathbf{6}})$            & 0.58~  & 0.51~  & 0.45~  & 0.40~  & 0.35~   \\ 
\hline
$\Gamma_{\Xi_{bb}}([^3P_2]_{\mathbf{6}})$            & 0.18~  & 0.16~  & 0.14~  & 0.12~  & 0.11~   \\
\hline
\end{tabular}
\label{mb}
\end{center}
\end{table}

Secondly in the pQCD calculation, the different selection of renormalization scale $\mu_r$ would also bring theoretical uncertainty. However, the decay width of this process is proportional to the strong coulping $\alpha^2_s(\mu_r)$ and we could easily analyze the scale uncertainty by calculating the strong coupling constant. Here we show the theoretical uncertainty caused by the renormalization scale by substituting three different scales, i.e., $\mu_r=2m_c$, $\rm{M}_{\it bc}$ or $2m_b$, in Table~\ref{mqun}. From Table~\ref{mqun}, we can see that all the results decrease as the scale increases.

\begin{table}[htb]
\begin{center}
\caption{Theoretical uncertainty caused by the renormalization scale for the production of $P$-wave $\Xi_{QQ^{\prime}}$ by substituting $\mu_r=2m_c$, $\rm{M}_{\it bc}$ or $2m_b$. The units of the decay widths are $10^{-9}$~GeV.}
\begin{tabular}{|c|c|c|c|} 
			\hline
			$\mu_r$(GeV)                                           & $2m_c$~  & $M_{bc}$~  & $2m_b$~     \\ 
			\hline\hline
			$\Gamma_{\Xi_{cc}}([^1P_1]_{\overline{\mathbf{3}}})$ & 1.08~  & 0.72~  & 0.60~  \\ 
			\hline
			$\Gamma_{\Xi_{cc}}([^3P_0]_{\mathbf{6}})$            & 0.83~  & 0.55~  & 0.46~  \\ 
			\hline
			$\Gamma_{\Xi_{cc}}([^3P_1]_{\mathbf{6}})$            & 0.86~  & 0.57~  & 0.47~  \\ 
			\hline
			$\Gamma_{\Xi_{cc}}([^3P_2]_{\mathbf{6}})$            & 0.29~  & 0.20~  & 0.16~  \\
			\hline\hline
			$\Gamma_{\Xi_{bc}}([^1P_1]_{\overline{\mathbf{3}}})$ & 6.60~  & 4.42~  & 3.65~  \\ 
			\hline
			$\Gamma_{\Xi_{bc}}([^1P_1]_{\mathbf{6}})$            & 3.30~  & 2.21~  & 1.83~  \\ 
			\hline
			$\Gamma_{\Xi_{bc}}([^3P_0]_{\overline{\mathbf{3}}})$ & 4.07~  & 2.72~  & 2.25~  \\ 
			\hline
			$\Gamma_{\Xi_{bc}}([^3P_0]_{\mathbf{6}})$            & 2.03~  & 1.36~  & 1.13~  \\ 
			\hline
			$\Gamma_{\Xi_{bc}}([^3P_1]_{\overline{\mathbf{3}}})$ & 16.07~ & 10.76~ & 8.89~  \\ 
			\hline
			$\Gamma_{\Xi_{bc}}([^3P_1]_{\mathbf{6}})$            & 8.04~  & 5.38~  & 4.45~  \\ 
			\hline
			$\Gamma_{\Xi_{bc}}([^3P_2]_{\overline{\mathbf{3}}})$ & 6.95~  & 4.65~  & 3.84~  \\ 
			\hline
			$\Gamma_{\Xi_{bc}}([^3P_2]_{\mathbf{6}})$            & 3.47~  & 2.33~  & 1.92~  \\ 
			\hline\hline
			$\Gamma_{\Xi_{bb}}([^1P_1]_{\overline{\mathbf{3}}})$ & 0.85~  & 0.57~  & 0.47~  \\ 
			\hline
			$\Gamma_{\Xi_{bb}}([^3P_0]_{\mathbf{6}})$            & 0.79~  & 0.53~  & 0.44~  \\ 
			\hline
			$\Gamma_{\Xi_{bb}}([^3P_1]_{\mathbf{6}})$            & 0.81~  & 0.54~  & 0.45~  \\ 
			\hline
			$\Gamma_{\Xi_{bb}}([^3P_2]_{\mathbf{6}})$            & 0.25~  & 0.17~  & 0.14~  \\
			\hline
		\end{tabular}
\label{mqun}
\end{center}
\end{table}

Finally, we shall discuss the theoretical uncertainty caused by the transition probability. Because of QCD confinement, the potential of diquark state in color sextuplet $\mathbf{6}$ is not exactly hydrogen-like. So there is a large uncertainty caused by the transition probability $h_{\mathbf{6}}$. It has been pointed out in literature~\cite{Zheng:2015ixa,Ma:2003zk} that $h_\mathbf{\bar{3}}$ is larger than $h_{\mathbf{6}}$ because of the ``one-gluon-exchange" interaction inside the diquark. The contribution from $h_{\mathbf{6}}$ shall be at least $v^2_r$-suppressed compared to $h_\mathbf{\bar{3}}$, and even can be ignored. When the transition probability $h_{\mathbf{6}}$ was taken as $h_{\mathbf{6}}/v^2_r \simeq h_\mathbf{\bar{3}}=|\Psi_{QQ'}'(0)|^2$ with $v^2_r=(0.1-0.3)$, there would be about $(2.76-3.05)\times 10^{3}$ $\Xi_{cc}$, $(4.45-4.87)\times 10^{4}$ $\Xi_{bc}$, $(1.73-1.95) \times 10^{3}$ $\Xi_{bb}$ events produced at the HL-LHC, and
$(0.17-0.18)\times 10^{2}$ $\Xi_{cc}$, $(2.69-2.82)\times 10^{2}$ $\Xi_{bc}$, $(0.10-0.11) \times 10^{2}$ $\Xi_{bb}$ events produced at the CEPC/ILC per year.
If the transition probability $h_{\mathbf{6}}=0$, there are still $2.61 \times 10^{3}$ $\Xi_{cc}$, $4.23 \times 10^{4}$ $\Xi_{bc}$, $1.61 \times 10^{3}$ $\Xi_{bb}$ events produced at the HL-LHC, and $0.16 \times 10^{2}$ $\Xi_{cc}$, $2.57 \times 10^{2}$ $\Xi_{bc}$, $0.10 \times 10^{2}$ $\Xi_{bb}$ events produced at the CEPC/ILC in one operation year.

\section{Summary}

Through the decay channel of Higgs coupled to the heavy quark, the indirect production mechanism of excited doubly heavy baryons was comprehensively analyzed based on the NRQCD framework via the process $H(p_0) \rightarrow \langle QQ'\rangle[n] \rightarrow \Xi_{QQ^{\prime}} (p_1)+ \bar {Q^{\prime}} (p_2) + \bar{Q} (p_3)$, where $Q$ and $Q^{\prime}$ denote as the heavy $c$ or $b$ quark for the production of $\Xi_{cc}$, $\Xi_{bc}$ and $\Xi_{bb}$. The intermediate $P$-wave diquark states include $\langle cc\rangle[^{1}P_{1}]_{\mathbf{\bar 3}}$, $\langle cc\rangle[^{3}P_{J}]_{\mathbf{6}}$, 
$\langle bc\rangle[^{1}P_{1}]_{\mathbf{\bar 3}/ \mathbf{6}}$, $\langle bc\rangle[^{3}P_{J}]_{\mathbf{\bar 3}/ \mathbf{6}}$, 
$\langle bb \rangle[^{1}P_{1}]_{\mathbf{\bar 3}}$ and $\langle bb\rangle[^{3}P_{J}]_{\mathbf{6}}$, with $J=0,1,2$. The results show that the decay widths of all summed $P$-wave $\Xi_{cc}$, $\Xi_{bc}$ and $\Xi_{bb}$ is about 3.05\%, 3.23\%, and 2.19\% of $S$-wave contributions. After further considering the contributions from $P$-wave states, the total decay widths and events expected to be produced per year at the HL-LHC through the corresponding process would be
\begin{eqnarray}
&&\Gamma_{H \rightarrow \Xi_{cc}+ \bar {c} + \bar {c}}=103.18^{+23.04}_{-16.04} \times 10^{-9}~\rm{GeV},~~~~N_{\Xi_{cc}}=0.41 \times 10^{4}, \nonumber\\
&&\Gamma_{H \rightarrow \Xi_{bc}+ \bar {c} + \bar {b}}=1616.63^{+1314.29}_{-632.45} \times 10^{-9}~\rm{GeV},~~~N_{\Xi_{bc}}=6.35 \times 10^{4}, \nonumber\\
&&\Gamma_{H \rightarrow \Xi_{bb}+ \bar {b} + \bar {b}}=69.94^{+7.32}_{-6.21} \times 10^{-9}~\rm{GeV},~~~~~~N_{\Xi_{bb}}=0.28 \times 10^{4}, \nonumber
\end{eqnarray}
where the theoretical uncertainty in decay widths comes from the mass of heavy quarks with a region $m_c=1.8\pm0.3$~GeV and $m_b=5.1\pm0.4$~GeV. And at HE-LHC, there would be more doubly heavy baryons produced than HL-LHC.
While at CEPC/ILC, a smaller number of events may be produced, only about $0.25\times 10^{2}$ $\Xi_{cc}$, $3.85\times 10^{2}$ $\Xi_{bc}$ and $0.17\times 10^{2}$ $\Xi_{bb}$. Although the produced events are small, the background of $e^{+}e^{-}$ colliders is cleaner to be measured by the experiments.
Considering a upgrade CEPC/ILC with the luminosity up to the level of HL-LHC, such as $3~ab^{-1}$, the produced events of corresponding $\Xi_{QQ'}$ would increase about 3.75 times.

From the invariant mass and angular differential distributions, we can find that 
the two antiquarks in the final state always move back to back. The differential distribution $d\Gamma/ d\rm s_{12}$ could reach a maximum when the invariant mass $\rm s_{12}$ is at its minimum value.
For the production of $P$-wave $\Xi_{cc}$ and $\Xi_{bb}$, all the kinematic features are the same, except that the differential decay width of $\Xi_{cc}$ is slightly larger than that of $\Xi_{bb}$. While for the excited $\Xi_{bc}$ production, the behaviors of the diquark states $\langle bc\rangle[^{3}P_{0}]_{\mathbf{\bar{3}/6}}$ change more dramatically than that of other $\langle bc\rangle$ diquark states in the differential distributions. And the differential decay widths could achieve its biggest when the $P$-wave $\Xi_{bc}$ move in the same direction as $\bar{c}$ and in the opposite direction as $\bar{b}$.

\black
\hspace{2cm}

{\bf Acknowledgements}: This work was partially supported by the Central Government Guidance Funds for Local Scientific and Technological Development, China (No. Guike ZY22096024) and the Guangxi Technology Base and Talent Subject (No. Guike AD20238014). This work was also supported by the National Natural Science Foundation of China (Grants No.12047506 and No.12005045).

\bibliographystyle{unsrt}

\end{document}